\documentclass[aps,floatfix,amsmath,nofootinbib,amssymb,superscriptaddress]{revtex4}

\usepackage{overpic}
\usepackage{amssymb}
\usepackage{indentfirst}
\usepackage{feynmf}   
\usepackage{slashed}  
\usepackage{cases}
\usepackage{color}
\usepackage{multirow}
\usepackage{epstopdf}
\usepackage{graphicx,color,bm}
\usepackage{epstopdf}

\usepackage[colorlinks,
            citecolor=blue,
            anchorcolor=red,
            menucolor=red,
            linkcolor=red,
            filecolor=red,
            runcolor=red,
            urlcolor=blue,
            frenchlinks=red]{hyperref}

\begin{document}

\title{The Collins asymmetry in electroproduction of Kaon at the electron ion colliders within TMD factorization}

\author{Shi-Chen Xue}
\affiliation{School of Physics and Microelectronics, Zhengzhou University, Zhengzhou, Henan 450001, China}
\author{Xiaoyu Wang}
\email{xiaoyuwang@zzu.edu.cn(Corresponding author)}
\affiliation{School of Physics and Microelectronics, Zhengzhou University, Zhengzhou, Henan 450001, China}
\author{De-Min Li}
\email{lidm@zzu.edu.cn}
\affiliation{School of Physics and Microelectronics, Zhengzhou University, Zhengzhou, Henan 450001, China}
\author{Zhun Lu}
\email{zhunlu@seu.edu.cn}
\affiliation{School of Physics, Southeast University, Nanjing 211189, China}

\begin{abstract}
We apply the transverse momentum dependent factorization formalism to investigate the transverse single spin dependent Collins asymmetry with a $\sin(\phi_h+\phi_s)$ modulation in the semi-inclusive production of Kaon meson in deep inelastic scattering process.
The asymmetry is contributed by the convolutions of the transversity distribution function $h_1(x)$ of the target proton and the Collins function of the Kaon in the final state. We adopt the available parametrization of $h_1(x)$ as well as the recent extracted result for the Kaon Collins function.
To perform the transverse momentum dependent evolution, the parametrization of the nonperturbative Sudakov form factor of the proton and final state Kaon are utilized.
We numerically predict the Collins asymmetry for charged
Kaon production at the electron ion colliders within the accuracy of next-to-leading-logarithmic order.
It is found that the asymmetry is sizable and could be measured.
We emphasize the importance of planned electron ion colliders in the aspect of constraining sea quark distribution functions as well as accessing the information of the nucleon spin structure and the hadronization mechanism.

\end{abstract}

\maketitle

\section{Introduction}

Understanding the 3-dimensional partonic structure of the spin-1/2 nucleon has been an active subject in Quantum Chromodynamics (QCD) spin physics and hadronic physics.
Various spin asymmetries in high energy scattering processes, such as the semi-inclusive deep inelastic scattering (SIDIS)~\cite{Collins:1992kk}, Drell-Yan~\cite{Jaffe:1991ra}, and electron-positron annihilation processes, have been recognized as useful tools to explore the internal structure of the nucleon.
In the kinematic region where the measured transverse momentum $P_{hT}$ of the produced hadron is much smaller than the invariant mass $Q$ of the virtual photon ($P_{hT}\ll Q$),
a convenient theoretical approach to study these processes is the transverse momentum dependent (TMD) factorization formalism, in which the differential cross section of SIDIS may be expressed as the convolution of the hard scattering factors, the TMD parton distribution functions (PDFs), and TMD fragmentation functions (FFs).
The transversity $h_1(x)$ is one of the eight TMD PDFs encoding the partonic structure of hadrons at leading-twist level~\cite{Bacchetta:2006tn}.
It describes the transverse polarization of the quark inside a transversely polarized nucleon, thereby it is one of the fundamental observables manifesting the nucleon structure.
However, compared to the unpolarized distribution and the helicity distribution, which have been extensively studied and measured,
the transversity is difficult to measure in high energy scattering process due to its odd chirality~\cite{Jaffe:1991kp}.
Another chiral-odd function is needed to couple with $h_1$ to ensure the chirality conservation.
Thus, there must be two hadrons participating the scattering process to insure the chirality has been flipped twice.
Either the two hadrons are both in the initial state~(Drell-Yan process) or one in the initial state and the other one in the final state~(SIDIS process).

In Drell-Yan process, the transversity can be accessed through the double transverse spin dependent asymmetry, which is contributed by the convolution of the quark transveresity and the antiquark transversity.
In SIDIS process under the TMD factorization, the chiral-odd probe to access transversity is the Collins function $H_1^\perp$~\cite{Collins:1992kk},
which describes the fragmenting of a transversely polarized quark to an unpolarized hadron.
The corresponding observable is the Collins asymmetry with a $\sin(\phi_h+\phi_S)$ modulation,
where $\phi_h$ and $\phi_s$ are the azimuthal angles for the transverse momentum of the outgoing hadron and the transverse spin of the nucleon target, respectively.
The Collins asymmetry in SIDIS process has been measured
by the HERMES Collaboration~\cite{Airapetian:2004tw,Airapetian:2009ae,Airapetian:2010ds,Airapetian:2012yg}, the COMPASS Collaboration~\cite{Adolph:2012sn}, and the JLab HALL A Collaboration~\cite{Qian:2011py,Zhao:2014qvx}.
The data obtained in SIDIS process combined with the one from $e^+e^-$ annihilation process can be utilized to simultaneously extract the valence quark transversity distribution function and Collins function~\cite{Anselmino:2007fs,Anselmino:2008jk,Anselmino:2013vqa,Anselmino:2015sxa,Anselmino:2015fty}.
Within the collinear factorization framework, the chiral-odd dihadron fragmentation function may be another promising probe in the dihadron production SIDIS process, which can also be applied to extract the valence quark transversity by combining the $e^+e^-$ annihilation data~\cite{Bacchetta:2011ip}.
Meanwhile,  the twist-3 collinear fragmentation function $\tilde{H}(z)$ may be used as the future probe~\cite{Wang:2016tix} to access the valence quark transversity through the $\sin\phi_S$ asymmetry in single transversely polarized SIDIS process, in which the transverse momentum of the final state hadron is integrated out.
Although much progress has been achieved, the information for the sea quark transversity distribution function is almost unknown due to the lack of the experimental data. The Kaon production SIDIS process may be an ideal analyzer to study the sea quark distribution function due to the strange constitute of the Kaon meson. Thereby, through the $\sin(\phi_h + \phi_S)$ Collins asymmetry in the Kaon meson production, there might be an opportunity to obtain the information of the sea quark transversity.

The purpose of this work is to evaluate the Collins asymmetry in Kaon production SIDIS process
at the kinematics region of the planned electron ion colliders (EICs), such as the proposed EIC in US~\cite{Accardi:2012qut} and the EIC in China (EicC)~\cite{Cao:2020}, in which the transversely polarized proton target will be available.
Since it is supposed that high luminosity can be realized at the EICs, the sea quark content of the nucleon may be explored with unprecedent accuracy.
The theoretical tool adopted in this study is the TMD factorization formalism~\cite{Collins:1981uk,Collins:1984kg,Collins:2011zzd,Ji:2004xq}, which has been widely applied to various high energy processes,
such as SIDIS~\cite{Collins:1981uk,Collins:2011zzd,Ji:2004wu,Aybat:2011zv,Collins:2012uy,Echevarria:2012pw},
$e^+ e^-$ annihilation~\cite{Collins:2011zzd,Pitonyak:2013dsu,Boer:2008fr}, Drell-Yan~\cite{Collins:2011zzd,Arnold:2008kf}, and W/Z production in hadron collision~\cite{Collins:2011zzd,Collins:1984kg,Lambertsen:2016wgj}.
In this framework, the differential cross section can be written as the convolution of the well-defined TMD PDFs and/or FFs.
The energy dependence of the TMD PDFs and FFs is encoded in the TMD evolution functions, the solution of which is usually given in $b$ space, which is conjugate to the transverse momentum space~\cite{Collins:1984kg,Collins:2011zzd} through Fourier transformation.
After solving the TMD evolution equation, the scale dependence of the TMDs may be included in the exponential form of the so-called Sudakov-like form factor~\cite{Collins:1984kg,Collins:2011zzd,Aybat:2011zv,Collins:1999dz}.
The Sudakov form factor can be further separated into perturbatively calculable part and the nonperturbative part, the latter one can not be calculated through perturbative theory and may be obtained by parameterizing experimental data.
We will investigate the effect of nonperturbative Sudakov form factor on the corresponding TMD PDFs and FFs when performing the TMD evolution to obtain the numerical results for the Collins asymmetry in Kaon production in SIDIS process.

The rest of the paper is organized as follows.
In Sec.~\ref{sec:formalisms}, we provide a detailed review on the Collins asymmetry in SIDIS process within the TMD factorization formalism.
Particularly, we present the procedure for the TMD evolution of the unpolarized and polarized TMDs involved in our calculations.
In Sec.~\ref{sec:numerical}, we perform the numerical estimate of the Collins asymmetry at the kinematics of two planned EICs.
We summarize the paper and discuss the results in Sec.~\ref{sec:conclusion}.

\section{The Collins asymmetry in SIDIS process within TMD factorization}
\label{sec:formalisms}

\begin{figure}
  \centering
  \includegraphics[width=0.4\columnwidth]{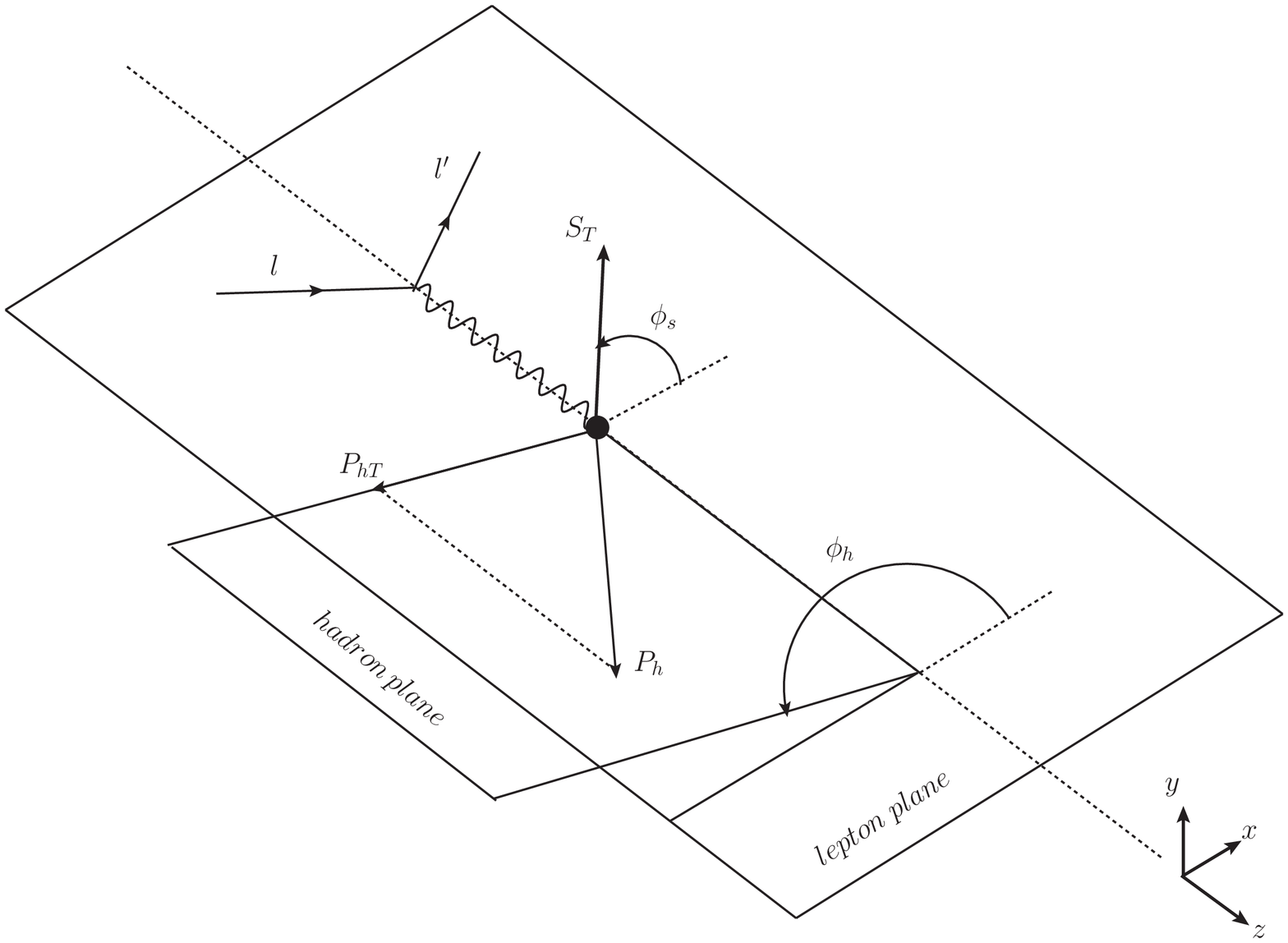}
  \caption{The definition of azimuthal angles in SIDIS~\cite{Bacchetta:2004jz}. The lepton plane is defined by $\ell$ and $\ell^\prime$. $S$ stands for the spin of the proton target with $S_T$ being the transverse component with respect to the virtual photon momentum. $P_{h}$ denotes the momentum of the produced Kaon. }
  \label{lhp}
\end{figure}

The process under study is the SIDIS process with Kaon production using an unpolarized electron beam scattering off a transversely polarized proton target
\begin{equation}
\label{eq:sidis}
e(\ell)+p^\uparrow(P) \longrightarrow e(\ell^\prime)+K(P_h)+X(P_X),
\end{equation}
where $\ell$ and $\ell^\prime$ stand for the four-momenta of the incoming and outgoing electrons, respectively, whereas $P$ and $P_h$ denote the four-momenta of the proton target and the final-state hadron~(which is Kaon in this work), respectively.
The reference frame of the studied SIDIS process is shown in Fig.~\ref{lhp}, in which the momentum direction of virtual photon is defined as $z-$axis according to the Trento conventions~\cite{Bacchetta:2004jz}. $P_{hT}$ and $S_T$ are the transverse component of $P_h$ and the spin vector $S$, respectively.
$\phi_h$ denotes the the azimuthal angle of the final hadron around the virtual photon, and $\phi_S$ stands for the angle between the lepton scattering plane and the direction of the transverse spin of the nucleon target.

We define the following invariants to express the differential cross section as well as the physical observables:
\begin{align}
&x_B=\frac {Q^2}{2P \cdot q}\,, \quad  y=\frac{P \cdot q}{P \cdot \ell}=\frac {Q^2}{x_B s}\,,\quad z_h=\frac{P \cdot P_h}{P \cdot q}\,,\nonumber \\
&Q^2=-q^2\,,\quad s=(P+\ell)^2\,, \nonumber
\end{align}
where $s$ is the total center of mass energy,
$q=\ell-\ell^\prime$ denotes the momentum of the virtual photon with invariant mass $Q^2$.
The 5-fold differential cross section~($x_B,y,z_h,\bm{P}_{hT}$) with a transversely polarized target has the following general form at twist-2 level~\cite{Kang:2015msa,Bacchetta:2006tn,Anselmino:2008sga},
\begin{align}
\frac{d^5\sigma(S_T)}{dx_B dy dz_h d^2\bm{P}_{hT}}
&= \sigma_0(x_B, y, Q^2)
\left[F_{UU} +    \sin(\phi_h+\phi_s)\, \frac{2 (1-y)}{1+(1-y)^2} \, F_{UT}^{\sin\left(\phi_h +\phi_s\right)} + \ldots\right],
  \label{eq:aut}
\end{align}
where $\sigma_0 = \frac{2\pi \alpha_{\rm em}^2}{Q^2}\frac{1+(1-y)^2}{y}$.
The ellipsis denotes other structure functions,
which we will not consider in this work.
Here, we only consider the Collins effect with the $\sin\left(\phi_h +\phi_s\right)$ modulation and neglect other structure functions.
$F_{UU}$ stands for the spin-averaged~(unpolarized) structure function and $F_{UT}^{\sin(\phi_h+\phi_s)}$ is the transverse spin-dependent structure function, which contributes by the coupling of the transversity distribution function of the proton target and the Collins function of the fragmenting of a transversely polarized quark to the final-state Kaon.

The Collins asymmetry with $\sin\left(\phi_h +\phi_s\right)$ modulation can be written in terms of $F_{UU}$ and $F_{UT}^{\sin(\phi_h+\phi_s)}$ as
\begin{eqnarray}
A_{UT}^{\sin(\phi_h+\phi_s)} = \frac{\sigma_0(x_B, y, Q^2)}{\sigma_0(x_B, y, Q^2)}  \frac{2 (1-y)}{1+(1-y)^2}
\frac{F_{UT}^{\sin\left(\phi_h +\phi_s\right)}}{F_{UU}},
\label{eq:asymmetry}
\end{eqnarray}
with $\frac{2 (1-y)}{1+(1-y)^2}$ the depolarization factor.
The structure functions $F_{UU}$ and $F_{UT}^{\sin(\phi_h+\phi_s)}$ in Eq.~(\ref{eq:asymmetry}) can be expressed as the convolution of the PDFs and FFs as~\cite{Bacchetta:2006tn}
\begin{align}
&F_{UU}(Q;P_{hT})=\mathcal{C}[f_1D_1], \label{fuu}\\
&F_{UT}^{\sin\left(\phi_h +\phi_s\right)}(Q;P_{hT}) = \mathcal{C}[\frac{-\hat{\bm h}\cdot \bm{k}_T}{M_h} h_1 H_1^\perp] . \label{fut}
\end{align}
Here, $f_1(x_B,\bm{p}_T)$ and $h_1(x_B,\bm{p}_T)$ are the unpolarized TMD distribution function and the transversity distribution function of the proton target, respectively. They depend on the Bjorken variable $x_B$ and the transverse momentum $\bm{p}_T$ of the quark inside the proton.
On the other hand, $D_1(z_h,\bm{k}_T)$ and $H_1^\perp(z_h,\bm{k}_T)$ are respectively the unpolarized fragmentation function and the Collins function, which depend on the longitudinal momentum fraction $z_h$ and the transverse momentum $\bm{k}_T$ of the final-state quark.
The notation $\mathcal{C}$ represents the convolution:
\begin{equation}
\label{eq:note_C}
\mathcal {C}\bigl[ \omega f  D \bigr]
= \sum_q e_q^2 \int d^2 \bm{p}_T  d^2 \bm{k}_T \delta^{(2)}\bigl(\bm{p}_T - \bm{k}_T - \bm{P}_{hT}/z_h \bigr)\omega(\bm{p}_T,\bm{k}_T)
f^q(x_B,p_T^2)\,D^q(z_h,k_T^2).
\end{equation}
Substituting Eq.~(\ref{eq:note_C}) into Eq.~(\ref{fuu}), one can obtain the unpolarized structure function as
\begin{align}
F_{UU}(Q;P_{hT})&=\mathcal{C}[f_1D_1] \nonumber\\
&=\sum_q e_q^2 \int  d^2 \bm{p}_T  d^2 \bm{k}_T \delta^{(2)}\bigl(\bm{p}_T - \bm{k}_T - \bm{P}_{h T}/z_h \bigr) f_1^q(x_B,p_T^2)\,D_1^q(z_h,k_T^2)\nonumber\\
&=\frac{1}{z_h^2}\sum_q e_q^2 \int  d^2 \bm{p}_T  d^2 \bm{K}_T \delta^{(2)}\bigl(\bm{p}_T + \bm{K}_T/z_h - \bm{P}_{h T}/z_h \bigr)f_1^q(x_B,p_T^2)\,D_1^q(z_h,K_T^2)\nonumber\\
&=\frac{1}{z_h^2}\sum_q e_q^2 \int  d^2 \bm{p}_T  d^2 \bm{K}_T \int \frac{d^2b}{(2\pi)^2}e^{-i\bigl(\bm{p}_T + \bm{K}_T/z_h - \bm{P}_{h T}/z_h \bigr)\cdot \bm{b}} f_1^q(x_B,p_T^2)\,D_1^q(z_h,K_T^2)\nonumber\\
&=\frac{1}{z_h^2}\sum_q e_q^2 \int \frac{d^2b}{(2\pi)^2}e^{i \bm{P}_{hT}/z_h \cdot \bm{b}} \tilde{f}_1^q(x_B,b)\,\tilde{D}_1^q(z_h,b).
\label{fuu_expanded}
\end{align}
Here, $\bm{K}_T =-z_h\bm{k}_T$ is the perpendicular momentum of the hadron $h$ with respect to the quark momentum, and we have applied the Fourier transformation to give the unpolarized TMD distributions and fragmentation functions in $b$ space (denoted by tildes):
\begin{align}
\int d^2 \bm{p}_T e^{-i\bm{p}_T \cdot \bm{b}}  f_1^q(x_B,p_T^2)&=\tilde{f}_1^q(x_B,b)\\
\int d^2 \bm{K}_T e^{-i \bm{K}_T/z_h \cdot \bm{b}} D_1^q(z_h,K_T^2)&=\tilde{D}_1^q(z_h,b)
\end{align}
We should point out that the energy dependence of the distribution functions and fragmentation functions has been neglected in Eq.~(\ref{fuu_expanded}) and will be discussed in detail in the following.

Similarly, the spin-dependent structure function $F_{UT}^{\sin\left(\phi_h +\phi_s\right)}$ in Eq.~(\ref{fut}) can be rewritten as
\begin{align}
F_{UT}^{\sin\left(\phi_h +\phi_s\right)}(Q;P_{hT}) &= \mathcal{C}[\frac{-\hat{\bm h}\cdot \bm{k}_T}{M_h} h_1 H_1^\perp]\nonumber\\
&=\sum_q e_q^2 \int  d^2 \bm{p}_T  d^2 \bm{k}_T \delta^{(2)}\bigl(\bm{p}_T - \bm{k}_T - \bm{P}_{h T}/z_h \bigr) \frac{-\hat{\bm h}\cdot \bm{k}_T}{M_h} h^q_1(x_B,p_T^2) H_1^{\perp,q}(z_h,k_T^2)\nonumber\\
&=\sum_q e_q^2 \frac{1}{z_h^2}\int  d^2 \bm{p}_T  d^2 \bm{K}_T \delta^{(2)}\bigl(\bm{p}_T +\bm{K}_T/z_h - \bm{P}_{h T}/z_h \bigr) \frac{\hat{\bm h}\cdot \bm{K}_T}{z_h M_h} h^q_1(x_B,p_T^2) H_1^{\perp,q}(z_h,K_T^2)\nonumber\\
&=\sum_q e_q^2 \frac{1}{z_h^2}\int  d^2 \bm{p}_T  d^2 \bm{K}_T \int \frac{d^2b}{(2\pi)^2}e^{-i\bigl(\bm{p}_T +\bm{K}_T/z_h - \bm{P}_{h T}/z_h \bigr)\cdot\bm{b}} \frac{\hat{\bm h}\cdot \bm{K}_T}{z_h M_h} h^q_1(x_B,p_T^2) H_1^{\perp,q}(z_h,K_T^2)\nonumber\\
&=\sum_q e_q^2 \frac{1}{z_h^2}\frac{1}{z_h} \int \frac{d^2b}{(2\pi)^2}e^{i\bm{P}_{h T}/z_h \cdot\bm{b}} \hat{ h}_\alpha \tilde h^q_1(x_B,b) \tilde H_1^{\perp \alpha,q}(z_h,b).
\label{fut_expanded}
\end{align}
Here, the transversity distribution function and Collins function in $b$ space are defined as
\begin{align}
\int d^2 \bm{p}_T e^{-i\bm{p}_T \cdot \bm{b}}  h_1^q(x_B,p_T^2)&=\tilde{h}_1^q(x_B,b);\\
 \int d^2 \bm{K}_T e^{-i \bm{K}_T/z_h \cdot \bm{b}} \frac{K_T^\alpha}{M_h}H_1^{\perp,q}(z_h,K_T^2)&=\tilde{H}_1^{\perp\alpha,q}(z_h,b).
\end{align}

\subsection{TMD evolution formalism}
The evolution of the TMDs is usually performed in $b$ space~(which is conjugated to the transverse momentum space~\cite{Collins:1984kg,Collins:2011zzd} through Fourier Transformation), since the cross section in $b$ space can be expressed as the simple production instead of the complicate convolution of functions in transverse momentum space.
After performing a reverse Fourier transformation from the $b$ space back to the transverse momentum space, we can obtain the TMD physical observables that can be measured experimentally.
Thus it is important to understand the $b$-behavior of the TMD functions.

Specifically, in TMD factorization the distributions $\tilde{F}(x_B,b;\mu,\zeta_F)$ and fragmentation functions $\tilde{D}(z_h,b;\mu,\zeta_D)$ in $b$ space have two kinds of energy dependence.
The first one is $\mu$ that is the renormalization scale related to the corresponding collinear PDFs/FFs,
and the other one is $\zeta_F~(\zeta_D)$ which is the energy scale serving as a cutoff to regularize the light-cone singularity in the operator definition of the TMDs.
The $\mu$ and $\zeta_F~(\zeta_D)$ dependences are encoded in different TMD evolution equations.
The energy evolution for the $\zeta_F~(\zeta_D)$ dependence of the TMD distributions~(fragmentation functions) is given by the Collins-Soper~(CS)~\cite{Collins:1984kg,Idilbi:2004vb,Collins:2011zzd} equation
\begin{align}
\frac{\partial\ \mathrm{ln} \tilde{F}(x_B,b;\mu,\zeta_F)}{\partial\ \ln \sqrt{\zeta_F}}=\frac{\partial\ \mathrm{ln} \tilde{D}(z_h,b;\mu,\zeta_D)}{\partial\ \mathrm{ln} \sqrt{\zeta_D}}=\tilde{K}(b;\mu),
\end{align}
while the $\mu$ dependence is given by the renormalization group equation
\begin{align}
&\frac{d\ \tilde{K}}{d\ \mathrm{ln}\mu}=-\gamma_K(\alpha_s(\mu)),\\
&\frac{d\ \mathrm{ln} \tilde{F}(x_B,b;\mu,\zeta_F)}
{d\ \mathrm{ln}\mu}=\gamma_F(\alpha_s(\mu);{\frac{\zeta^2_F}{\mu^2}}),\\
&\frac{d\ \mathrm{ln} \tilde{D}(z_h,b;\mu,\zeta_D)}
{d\ \mathrm{ln}\mu}=\gamma_D(\alpha_s(\mu);{\frac{\zeta^2_D}{\mu^2}}),
\end{align}
with $\alpha_s$ the strong coupling at the energy scale $\mu$, $\tilde{K}$ the CS evolution kernel, and $\gamma_K$, $\gamma_F$ and $\gamma_D$ the anomalous dimensions.
Hereafter, we will set $\mu=\sqrt{\zeta_F}=\sqrt{\zeta_D}=Q$, and express respectivley $\tilde{F}(x_B,b;\mu=Q,\zeta_F=Q^2)$ and $\tilde{D}(z_h,b;\mu=Q,\zeta_F=Q^2)$ as $\tilde{F}(x_B,b;Q)$ and $\tilde{D}(z_h,b;Q)$ for simplicity.

Solving these TMD evolution equations, one can obtain the solution of the energy dependence for TMDs, which has the general form as
\begin{equation}
\tilde{F}(x_B,b;Q)=\mathcal{F}\times e^{-S}\times \tilde{F}(x_B,b;\mu),
\label{eq:f}
\end{equation}
where $\mathcal{F}$ is the factor related to the hard scattering, $S$ is the Sudakov-like form factor. Eq.~(\ref{eq:f}) shows that the energy evolution of TMD distributions from an initial energy $\mu$ to another energy $Q$ is encoded in the Sudakov-like form factor $S$ by the exponential form $\mathrm{exp}(-S)$.
Similarly, for fragmentation functions, they have the same solution structure as
\begin{equation}
\tilde{D}(z_h,b;Q)=\mathcal{D}\times e^{-S}\times \tilde{D}(z_h,b;\mu),
\label{eq:D}
\end{equation}
where $\mathcal{D}$ is the factor related to the hard scattering.
The coefficients $\mathcal{F}$ and $\mathcal{D}$ depend on the factorization schemes, which have been studied in details in Ref~\cite{Prokudin:2015ysa}.

In the small $b$ region ($b\ll1/\Lambda_{\textrm{QCD}}$), the $b$ dependence of TMDs is perturbative and can be calculated by QCD.
However, the dependence in large $b$ region turns to non-perturbative, since the operators are separated by a large distance.
It is convenient to include a nonperturbative Sudakov-like form factor $S_{\rm NP}$ to take into account the TMD evolution effect in the large $b$ region, and the form factor which is usually given in a parametrization form and must be obtained by analyzing experimental data, given the present lack of non-perturbative calculations.
To combine the perturbative information at small $b$ with the nonperturbative part at large $b$,
a matching procedure should be introduced with a parameter $b_{\mathrm{max}}$ serving as the boundary between the two regions.
A $b$-dependent function $b_\ast$ is defined to have the property $b_\ast\approx b$ in small $b$ region and $b_{\ast}\approx b_{\mathrm{max}}$ in large $b$ region,
\begin{align}
b_\ast=\frac{b}{\sqrt{1+b^2/b_{\rm max}^2}} \,  \, , \,  \, b_{\rm max}< 1/\Lambda_{\textrm{QCD}}  \,,
\end{align}
which was introduced in the original CSS prescription~\cite{Collins:1984kg}.
The prescription also allows for a smooth transition from perturbative to nonperturbative regions and avoids the Landau pole singularity in $\alpha_s(\mu_b)$.
The typical value of $b_{\mathrm{max}}$ is chosen around $1\ \mathrm{GeV}^{-1}$ to guarantee that $b_{\ast}$ is always in the perturbative region.

In the small $b$ region, the TMDs can be expressed as the convolutions of the perturbatively calculable hard coefficients and the corresponding collinear counterparts at fixed energy $\mu$, which could be the collinear PDFs/FFs or the multiparton correlation functions~\cite{Collins:1981uk,Bacchetta:2013pqa}
\begin{equation}
\tilde{F}_{q/H}(x,b;\mu)=\sum_i C_{q\leftarrow i}\otimes F_{i/H}(x,\mu),
\label{eq:small_b_F}
\end{equation}
where $\otimes$ stands for the convolution in the momentum fraction $x$,
\begin{align}
 C_{q\leftarrow i}\otimes F^{i/H}(x_B,\mu_b)& \equiv \int_{x_B}^1\frac{dx}{x} C_{q\leftarrow i}(\frac{x_B}{x},b;\mu_b)f^{i/H}(x,\mu_b), \\
 \hat C_{j\leftarrow q} \otimes D^{H/j}(z_h,\mu_b)& \equiv \int_{z_h}^1\frac{dz}{z} \hat C_{j\leftarrow q}(\frac{z_h}{z},b;\mu_b)D^{H/j}(z,\mu_b),
 \label{eq:otimes}
\end{align}
and $F_{i/H}(\xi,\mu)$ is the corresponding collinear counterpart of flavor $i$ in hadron $H$ at the energy scale $\mu$, which could be a dynamic scale related to $b_*$ by $\mu_b=c_0/b_*$, with $c_0=2e^{-\gamma_E}$ and the Euler constant $\gamma_E\approx0.577$~\cite{Collins:1981uk}.
In addition, the sum $\Sigma i$ runs over all parton flavors.
Independent on the type of initial hadrons, the perturbative hard coefficients $C_{q\leftarrow i}$ have been calculated for the parton-target case~\cite{Collins:1981uw,Aybat:2011zv} as the series of $(\alpha_s/\pi)$ and the results have been presented in Ref.~\cite{Bacchetta:2013pqa}~(see also Appendix A of Ref.~\cite{Aybat:2011zv}).
Thus, the general expression of TMDs in $b$ space in Eqs.~(\ref{eq:f}),~(\ref{eq:D}) can be rewritten as
\begin{equation}
\tilde{F}(x_B,b;Q)=\mathcal{F}\times e^{-S}\times \sum_i C_{q\leftarrow i}\otimes F_{i/H}(x_B,\mu).
\label{eq:f_fixed_evo}
\end{equation}

\begin{equation}
\tilde{D}(z_h,b;Q)=\mathcal{D}\times e^{-S} \times \sum_i \hat{C}_{q\leftarrow i}\otimes D_{H/i}(z_h,\mu),
\label{eq:D_fixed_evo}
\end{equation}

The Sudakov-like form factor $S$ can be separated into the perturbatively calculable part ${S}_{\rm pert}(Q;b_*)$ and the nonperturbative part $S_{\rm NP}(Q;b)$
\begin{equation}
\label{eq:S}
{S}(Q;b)= {S}_{\rm pert}(Q;b_*)+S_{\rm NP}(Q;b).
\end{equation}
According to the intensive studies in Refs.~\cite{Echevarria:2014xaa,Kang:2011mr,Aybat:2011ge,Echevarria:2012pw,Echevarria:2014rua},
the perturbative part of the Sudakov-like form factor ${ S}_{\rm pert}(Q;b_*)$ has the general form
\begin{equation}
\label{eq:Spert}
{ S}_{\rm pert}(Q;b_*)=\int^{Q^2}_{\mu_b^2}\frac{d\bar{\mu}^2}{\bar{\mu}^2}\left[A(\alpha_s(\bar{\mu}))
\ln(\frac{Q^2}{\bar{\mu}^2})+B(\alpha_s(\bar{\mu}))\right].
\end{equation}
For different kinds of TMDs, ${ S}_{\rm pert}(Q;b_*)$ is universal and has the same result, namely, ${ S}_{\rm pert}(Q;b_*)$ is spin independent.
The coefficients $A$ and $B$ in Eq.(\ref{eq:Spert}) can be expanded as the series of $\alpha_s/{\pi}$:
\begin{align}
A=\sum_{n=1}^{\infty}A^{(n)}(\frac{\alpha_s}{\pi})^n,\\
B=\sum_{n=1}^{\infty}B^{(n)}(\frac{\alpha_s}{\pi})^n.
\end{align}
Here, we list $A^{(n)}$ to $A^{(2)}$ and $B^{(n)}$ to $B^{(1)}$ up to the accuracy of next-to-leading-logarithmic (NLL) order~\cite{Collins:1984kg,Landry:2002ix,Qiu:2000ga,Kang:2011mr,Aybat:2011zv,Echevarria:2012pw}:
\begin{align}
A^{(1)}&=C_F,\\
A^{(2)}&=\frac{C_F}{2}\left[C_A\left(\frac{67}{18}-\frac{\pi^2}{6}\right)-\frac{10}{9}T_Rn_f\right],\\
B^{(1)}&=-\frac{3}{2}C_F.
\end{align}

However, the nonperturbative Sudakov-like form factor $S_{\rm NP}(Q;b)$ can not be obtained from perturbative calculation, and is usually parameterized from experimental data.
The general form of $S_{\rm NP}(Q;b)$ was suggested as~\cite{Collins:1984kg}
\begin{equation}
\label{eq:snp_gene}
S_{\rm NP}(Q;b)=g_2(b)\ln(\frac{Q}{Q_0})+g_1(b)\ .
\end{equation}
The functions $g_1(b)$ and $g_2(b)$ contain different information. $g_2(b)$ provides the evolution information of the CS kernel $\tilde K$ in large $b$ region, which does not depend on the particular process and has universal expression for all kinds of
TMDs~\cite{Collins:2011zzd,Aybat:2011zv,Echevarria:2014xaa,Kang:2015msa}.
It has no dependence on momentum fractions and energy scale.
Also, $g_2(b)$ shall follow the power behavior as $b^2$ at small-$b$ region, according to the power counting analysis in Ref.~\cite{Korchemsky:1994is}, which can be an essential constraint for the parametrization of $g_2(b)$.
There are several parameterizations for $S_{\rm NP}$ in literature.
The original BLNY fit parameterized $S_{\rm NP}$ in Drell-Yan process as~\cite{Landry:2002ix}
\begin{equation}
\label{snp:BLNY}
\left(g_1+g_2\ln (Q/2Q_0)+g_1g_3\ln(100\,x_1x_2)\right)b^2,
\end{equation}
where $x_1$ and $x_2$ are the longitudinal momentum fractions of the incoming hadrons carried by
the initial state quark and antiquark.
The BLNY-fit proved to be reliable in the description of the Drell-Yan data and $W^\pm,Z$ boson production~\cite{Landry:2002ix}.
However, the BLNY-type fit~\cite{Sun:2013dya,Su:2014wpa} can not describe the transverse momentum distribution of hadron when the parametrization was extrapolated to the typical SIDIS kinematics in HERMES and COMPASS.
Inspired by Refs.~\cite{Landry:2002ix,Konychev:2005iy}, a widely used parametrization of $S_{\rm NP}$ for TMD distributions or fragmentation functions was proposed~\cite{Landry:2002ix,Konychev:2005iy,Davies:1984sp,Ellis:1997sc,Bacchetta:2013pqa,Echevarria:2014xaa}
\begin{align}
\label{snp:gaussian}
S_{\rm NP}^{\rm pdf/ff} &= b^2\left(g_1^{\rm pdf/ff}+ \frac{g_2}{2} \ln\frac{Q}{Q_0}\right),
\end{align}
where the factor $1/2$ in front of $g_2$ comes from the fact that only one hadron is involved for the parametrization of $S_{\rm NP}^{\rm pdf/ff}$, while the parameter in Ref.~\cite{Konychev:2005iy} is for $pp$ collisions.
The parameter $g_1^{\rm pdf/ff}$ in Eq.~(\ref{snp:gaussian}) depends on the type of TMDs, which can be regarded as the width of the intrinsic transverse momentum for the relevant TMDs at the initial energy scale $Q_0$~\cite{Qiu:2000ga,Aybat:2011zv,Anselmino:2012aa}.
Assuming a Gaussian form for the dependence of the transverse momentum, one can obtain
\begin{align}
g_1^{\rm pdf} = \frac{\langle k_\perp^2\rangle_{Q_0}}{4},
\qquad
g_1^{\rm ff} = \frac{\langle p_\perp^2\rangle_{Q_0}}{4z_h^2},
\label{eq:g1}
\end{align}
where $\langle k_\perp^2\rangle_{Q_0}$ and $\langle p_\perp^2\rangle_{Q_0}$
represent the relevant averaged intrinsic transverse momenta squared for TMD distributions and TMD fragmentation functions at the initial scale $Q_0$, respectively.
It is shown in Ref.~\cite{Echevarria:2014xaa} that the form in Eq.~(\ref{snp:gaussian}) and a universal $g_2 =0.184$ can describe the SIDIS and Drell-Yan data.

To release the tension between the original BLNY fit to the Drell-Yan type data
and the fit to the SIDIS data from HERMES/COMPASS in the CSS resummation formalism, in Ref.~\cite{Su:2014wpa} the authors proposed a new form for $S_{\rm NP}$ in which
the $g_2(b)$ term was modified to the form of $\ln(b/b_\ast)$ and the functional form of $S_{\rm NP}$ turned to~\cite{Su:2014wpa}
\begin{equation}
g_1b^2+g_2\ln(b/b_*)\ln(Q/Q_0)+g_3b^2\left((x_0/x_1)^\lambda+(x_0/x_2)^\lambda\right)\ . \label{eq:siyy}
\end{equation}
This form has been suggested in an early study by Collins and Soper~\cite{Collins:1985xx}, but has not yet
been adopted in any phenomenological analysis until the calculation in Ref.~\cite{Su:2014wpa}.
The comparison between the original BLNY parametrization and this form shows that the new form of $S_{\rm NP}$ can fit the experimental data of Drell-Yan type process as equally well as the original BLNY parametrization.
In Ref.~\cite{Kang:2015msa}, the form in Eq.~(\ref{eq:siyy}) was adopted to simultaneously extract the valence transversity distributions and the pion Collins fragmentation functions, as well as the non-perturbative Sudakov form factors, from a global fit of the current experimental data on $e^+e^-$ annihilations measured by BELLE and BABAR collaborations and SIDIS data from HERMES, COMPASS, and JLab HALL A experiments.

In Ref.~\cite{Bacchetta:2017gcc}, the $g_2(b)$ function was parameterized as the Gaussian form $g_2 b^2$, following the BLNY convention.
Furthermore, in the function $g_1(b)$, the Gaussian width also depends on $x$.
The authors simultaneously fit the experimental data of SIDIS process from HERMES and COMPASS Collaborations, the Drell-Yan events at low energy, and the $Z$ boson production with totally 8059 data points. The extraction can describe the data well in the regions where TMD factorization is supposed to hold.
In Ref.~\cite{Bertone:2019nxa}, the unpolarized TMD PDFs and the non-perturbative part of TMD evolution kernel were extracted from the global analysis of Drell-Yan and Z-boson production data at the next-to-next-to-leading order (NNLO) in perturbative QCD.
The parametrization form applies a more flexible parametrization of $S_{\rm NP}$ with five free parameters, which is able to accommodate a range of different behaviors, such as the exponential and the Gaussian one.
In Ref.~\cite{Bacchetta:2019sam}, a flexible functional form of the non-perturbative contributions was adopted with nine free parameters, which turned out to be all well constrained, with moderate correlations amongst them. It shows explicit $x$ dependence, which is mostly constrained by the data at 7 and 8 TeV from ATLAS and can also demonstrates that most of the data sets are not sensitive to the $x$ dependence of TMDs.

As the information on the Sudakov form factor for the Kaon fragmentation functions is still unknown, we assume the Gaussian form for $g_2(b)$ in Eq.~(\ref{snp:gaussian}) to perform the TMD evolution for the distributions and fragmentation functions.
One can obtain the nonperturbative Sudakov form factor for the PDF and FF as
\begin{align}
S_{\rm NP}^{\rm pdf}(Q;b)
&=\frac{g_2}{2} \ln(\frac{Q}{Q_0})b^2 +g_1^{\rm pdf}b^2,\nonumber\\
S_{\rm NP}^{\rm ff}(Q;b)
&=\frac{g_2}{2} \ln(\frac{Q}{Q_0})b^2 +g_1^{\rm ff}b^2.
\label{eq:sud_np}
\end{align}

Combining all the parts together, the scale-dependent TMDs in $b$ space as functions of $x_B/z_h$, $b$, and $Q$ can be rewritten as
\begin{align}
\tilde{F}_{q/H}(x_B,b;Q)&=e^{-\frac{1}{2}S_{\mathrm{Pert}}(Q;b_\ast)
-S^{\rm pdf}_{\mathrm{NP}}(Q;b)}\mathcal{F}(\alpha_s(Q))
\sum_iC_{q\leftarrow i}\otimes F^{i/H}(x_B,\mu_b),  \label{eq:F_evo1} \\
\tilde{D}_{H/q}(z_h,b;Q)&=e^{-\frac{1}{2}S_{\mathrm{Pert}}(Q;b_\ast)
-S^{\rm ff}_{\mathrm{NP}}(Q;b)}\mathcal{D}(\alpha_s(Q))
\sum_j \hat C_{j\leftarrow q} \otimes D^{H/j}(z_h,\mu_b). \label{eq:D_evo1}
\end{align}
Performing a Fourier transformation from the $b$ space back to the $\bm{k}_\perp$ can lead to the TMD distribution functions and fragmentation functions.

\subsection{Unpolarized structure function}

In this subsection, we will turn to the expression of the unpolarized structure function $F_{UU}$ in Eq.~(\ref{fuu}) in terms of the unpolarized distribution function $f_1$ and the unpolarized fragmentation function $D_1$.
Using the general representation of the distribution function and fragmentation function in Eqs.~(\ref{eq:F_evo1})~and~(\ref{eq:D_evo1}),
the unpolarized distribution function $\tilde{f}_1(x_B,b,Q)$ and the unpolarized fragmentation function $\tilde{D}_1(z_h,b,Q)$ can be written as
\begin{align}
\tilde{f}_1^{q/p}(x_B,b;Q)&=e^{-\frac{1}{2}S_{\mathrm{Pert}}(Q;b_\ast)
-S^{f_1}_{\mathrm{NP}}(Q;b)}\mathcal{F}(\alpha_s(Q))
\sum_iC_{q\leftarrow i}\otimes f^{i/p}_1(x_B,\mu_b), \\
\tilde{D}_1^{H/q}(z_h,b;Q)&=e^{-\frac{1}{2}S_{\mathrm{Pert}}(Q;b_\ast)
-S^{D_1}_{\mathrm{NP}}(Q;b)}\mathcal{D}(\alpha_s(Q))
\sum_j \hat C_{j\leftarrow q} \otimes D_1^{H/j}(z_h,\mu_b).
\end{align}
Substituting them into the unpolarized structure function in Eq.~(\ref{fuu_expanded}), one can have the following expression
\begin{align}
&F_{UU}(Q;P_{hT})
=\frac{1}{z_h^2}\int \frac{d^2b}{(2\pi)^2} e^{i\bm{P}_{h T}/z_h \cdot\bm{b}} \widetilde {F}_{UU}(Q;b)
\end{align}
with the structure function in $b$-space as
\begin{align}
&\widetilde {F}_{UU}(Q;b)=e^{-{S}_{\rm pert}(Q;b_*)-S_{\rm NP}^{\rm SIDIS}(Q;b)}
\widetilde{F}_{UU}(b_*).
\end{align}
Here nonperturbative Sudakov form factor $S_{\rm NP}^{\rm SIDIS}(Q;b)$ is the combination of the unpolarized distribution function part and the unpolarized fragmentation function part
\begin{align}
S_{\rm NP}^{\rm SIDIS}(Q;b)
&= S_{\rm NP}^{\rm f_1}(Q;b)+S_{\rm NP}^{\rm D_1}(Q;b)  \nonumber\\
&=g_2 \ln(\frac{Q}{Q_0})b^2 +g_1^{\rm f_1}b^2+g_1^{\rm D_1}b^2,
\end{align}
where $g_1^{\rm f_1}$ and $g_1^{\rm D_1}$, given by Eq.~(\ref{eq:g1}), can be extracted from experimental data in SIDIS, Drell-Yan, and $e^+ e^-$ annihilation.
And the structure function in the perturbative $b$-region can be written as
\begin{align}
&\widetilde{F}_{UU}(b_*) = \sum_{q} e_q^2 \,\left(\mathcal{F}(\alpha_s(Q))
\sum_iC_{q\leftarrow i}\otimes f^{i/p}_1(x_B,\mu_b)\right)
\left(\mathcal{D}(\alpha_s(Q))
\sum_j \hat C_{j\leftarrow q} \otimes D_1^{H/j}(z_h,\mu_b)\right).
\end{align}
We would like to point out that the coefficients and factors related to the perturbative hard scattering above depend on the TMD factorization scheme used to define the TMD operators.
However, one can defined scheme-independent coefficients $C^{\rm (SIDIS)}$ which absorb all the scheme-dependent hard factors and coefficients.
The expressions of them can be found in Ref.~\cite{Kang:2015msa}:
\begin{eqnarray}
C_{q\gets q'}^{\rm (SIDIS)}(x,\mu_b)&=& \delta_{q'q} [\delta(1-x)+\frac{\alpha_s}{\pi}(\frac{C_F}{2}(1-x)  -2C_F\delta(1-x)  )]\; , \label{eq:cf_css}
\\
C_{q\gets g}^{\rm (SIDIS)}(x,\mu_b)&=& \frac{\alpha_s}{\pi} {T_R} \, x (1-x)\; ,  \label{eq:cf1_css}
\\
\hat C_{q'\gets q}^{\rm (SIDIS)}(z,\mu_b)&=& \delta_{q'q} [\delta(1-z)+\frac{\alpha_s}{\pi}(\frac{C_F}{2}(1-z)  -2C_F\delta(1-z) + P_{q\gets q}(z)\, \ln z) ]\; ,  \label{eq:cd_css}
\\
\hat C_{g\gets q}^{\rm (SIDIS)}(z,\mu_b)&=& \frac{\alpha_s}{\pi} \left( \frac{C_F}{2} z\; +  P_{g\gets q}(z)\, \ln z \right)
 \label{eq:cd1_css}
\end{eqnarray}
with the splitting functions $P_{q\gets q}$ and $P_{g\gets q}$ have the general form
\begin{align}
P_{q\gets q}(z) &= C_F \left[ \frac{1+z^2}{(1-z)_+} + \frac{3}{2} \delta(1-z) \right] \, ,
\label{P_qq}
\\
P_{g\gets q}(z) &= C_F \frac{1+(1-z)^2}{z} \; ,
\label{P_gq}
\end{align}
where $C_F=4/3$, $T_R=1/2$, and the subscript symbol ``$+$'' denotes ``$+$ function''.
Using the scheme-independent coefficients $C_{q\gets i}^{\rm (SIDIS)}$ and $\hat C_{j\gets q}^{\rm (SIDIS)}$, the $\widetilde{F}_{UU}(b_*)$  can be written as
\begin{eqnarray}
\widetilde{F}_{UU}(b_*) = \sum_{q} e_q^2 \, \left( \sum_i C_{q\gets i}^{\rm (SIDIS)}\otimes f_{1}^{i/p}(x_B,\mu_b) \right) \times\left(\sum_j \hat{C}_{j\gets q}^{\rm (SIDIS)}\otimes D_{1}^{K/j}(z_h,\mu_b)\right)  \ .
\label{eq:fuu1}
\end{eqnarray}
Therefore, the final result of the unpolarized structure function in Eq.~(\ref{fuu}) can be obtained as
\begin{align}
F_{UU}(Q;P_{hT})
&=\frac{1}{z_h^2}\int_0^\infty  \frac{db~b}{(2\pi)}J_0(P_{h T}/z_h~b) \widetilde {F}_{UU}(Q;b)\nonumber\\
&=\frac{1}{z_h^2}\sum_q e_q^2 \int_0^\infty  \frac{db~b}{(2\pi)}J_0(P_{h T}/z_h~b) e^{-{S}_{\rm pert}(Q;b_*)-S_{\rm NP}^{\rm SIDIS}(Q;b)}\nonumber\\
 &~~\left( \sum_i C_{q\gets i}^{\rm (SIDIS)}\otimes f_{1}^{i/p}(x_B,\mu_b) \right) \times\left(\sum_j \hat{C}_{j\gets q}^{\rm (SIDIS)}\otimes D_{1}^{K/j}(z_h,\mu_b)\right)\ .
\label{fuu_res}
\end{align}

\subsection{Spin-dependent structure function}

Using the general representation of the distribution function in Eq.~(\ref{eq:F_evo1}),
the proton transversity distribution in $b$ space $\tilde{h}_1(x_B,b,Q)$ can be written as
\begin{align}
\label{eq:h1}
\tilde{h}^{q/p}_{1}(x_B,b;Q)&=e^{-\frac{1}{2}S_{\mathrm{Pert}}(Q;b_\ast)
-S^{\rm h_1}_{\mathrm{NP}}(Q;b)}\mathcal{H}(\alpha_s(Q))
\sum_i\delta C_{q\leftarrow i}\otimes h_{1}^{i/p}(x_B,\mu_b),
\end{align}
where $\mathcal{H}(\alpha_s(Q))$ is the hard scattering factor, $\delta C_{q\leftarrow i}$ is the coefficient related to the transversity distribution, and $h_1^{i/p}(x_B,\mu_b)$ is the collinear transversity distribution of flavor $i$ in the proton at the energy scale $\mu_b$.
In the perpurbative $b$-region, the Collins function may be expressed as the convolution of the perturabtively calculable coefficients ${\delta \hat C}_{j\leftarrow q}$ and the corresponding twist-3 collinear correlation function $\hat{H}_{h/j}^{(3)}$:
\begin{align}
\label{eq:H1}
\tilde{H}^{\perp \alpha}_{1,h/q}(z_h,b;\mu_b)&= (\frac{ib^\alpha}{2})
\sum_j{\delta \hat C}_{j\leftarrow q}\otimes \hat{H}_{h/j}^{(3)}(z_h,\mu_b).
\end{align}
Furthermore,
$\hat{H}_{K/j}^{(3)}(z_h,\mu_b)$  is related to the first-$p_T$ momentum of the Collins function as~\cite{Kang:2015msa}
\begin{align}
\hat{H}_{h/j}^{(3)}(z_h)=\int d^2p_\perp \frac{|p_\perp^2|}{M_h} H_{1\, h/j}^\perp(z_h,p_\perp)\; .
\label{eq:H3}
\end{align}
Therefore, utilizing the general representation of the fragmentation function in Eq.~(\ref{eq:D_evo1}), the Collins function in $b$-space can be rewritten as
\begin{align}
\label{eq:Col}
\tilde{H}^{\perp \alpha}_{1,K/q}(z_h,b;Q)&=(\frac{ib^\alpha}{2})e^{-\frac{1}{2}S_{\mathrm{Pert}}(Q;b_\ast)
-S^{\rm H^{\perp}_1}_{\mathrm{NP}}(Q;b)}\mathcal{H}_{\rm Collins}(\alpha_s(Q))
\sum_j\delta \hat{C}_{j\leftarrow q}\otimes\hat{H}_{K/j}^{(3)}(z_h,\mu_b)
\end{align}
with $\mathcal{H}_{\rm Collins}(\alpha_s(Q))$ being the hard factor.

Substituting the transversity distribution function and the Collins function in Eqs.~(\ref{eq:h1}) and  (\ref{eq:Col}) into the spin-dependent structure function in Eq.~(\ref{fut_expanded}), one can obtain
\begin{align}
F_{UT}^{\sin\left(\phi_h +\phi_s\right)}(Q;P_{hT})
&= \frac{1}{z_h^2}\frac{1}{z_h} \int \frac{d^2b}{(2\pi)^2}e^{i\bm{P}_{h T}/z_h \cdot\bm{b}} \frac{q_{T,\alpha}}{\bm{q}_T} \widetilde {F}_{UT}^\alpha(Q;b),
\end{align}
where ${F}_{UT}^\alpha(Q;b)$ is the spin-dependent structure function in $b$-space and has the evolved form
\begin{eqnarray}
\widetilde {F}_{UT}^\alpha(Q;b)&=(\frac{i b^\alpha}{2})e^{-{S}_{\rm pert}(Q;b_*)-S_{\rm NP~Collins}^{\rm SIDIS}(Q;b)}
\widetilde{F}_{UT}(b_*).
\end{eqnarray}
The nonperturbative part of the Sudakov-like form factor receives contributions from the transversity distribution and the Collins function:
\begin{align}
S_{\rm NP\; collins}^{\rm SIDIS}(Q;b)
&= S_{\rm NP}^{\rm h_1}(Q;b)+S_{\rm NP}^{\rm H^{\perp}_1}(Q;b)  \nonumber\\
&=g_2 \ln\left(\frac{Q}{Q_0}\right)b^2 +g_1^{\rm h_1}b^2+g_1^{\rm H^{\perp}_1}b^2,
\end{align}
and $\widetilde{F}_{UT}(b_*)$ is the structure function in the perturbative $b_\ast$ region as
\begin{eqnarray}
\widetilde{F}_{\rm collins}(b_*) &=& \sum_{q} e_q^2 \, \left( \mathcal{H}(\alpha_s(Q))\sum_i \delta C_{q\gets i}\otimes h_{1}^{i/p}(x_B,\mu_b) \right)\left(\mathcal{H}_{\rm Collins}(\alpha_s(Q)) \sum_j \delta \hat{C}_{j\gets q}\otimes \hat H_{K/j}^{(3)}(z_h,\mu_b)\right)\, .
\label{eq:fcollins}
\end{eqnarray}
Similar to the unpolarized case, all of the hard scattering factor and coefficients can be absorbed in a new $C$-coefficients to remove the scheme dependence, which leads to the new $C$-coefficients in the spin-dependent case as
\begin{align}
&\delta C_{q\gets q'}^{\rm (SIDIS)}(z,\mu_b)  (x,\mu_b) = \delta_{q'q} \left[\delta(1-x)+\frac{\alpha_s}{\pi}(-2C_F\delta(1-x)) \right]\;,\nonumber\\
&\delta \hat C_{q'\gets q}^{\rm (SIDIS)}(z,\mu_b) = \delta_{q'q} [\delta(1-z)+\frac{\alpha_s}{\pi}(\hat P_{q\gets q}^{c}(z) \ln z  -2C_F\delta(1-z)) ]\; ,
\end{align}
where the function $\hat P_{q\gets q}^{c}(z)$ is given by
\begin{align}
\hat P_{q\gets q}^{c}(z) = C_F [ \frac{2 z}{(1- z)_+} +\frac{3}{2}\delta(1-z)].
\label{P_qqc}
\end{align}
Thus, $\widetilde{F}_{UT}(b_*)$ can be rewritten as
\begin{eqnarray}
\widetilde{F}_{\rm collins}(b_*) &=& \sum_{q} e_q^2 \, \left( \sum_i\delta C^{\rm (SIDIS)}_{q\gets i}\otimes h_{1}^{i/p}(x_B,\mu_b) \right)\left(\sum_j\delta \hat{C}_{j\gets q}^{\rm (SIDIS)}\otimes \hat H_{K/j}^{(3)}(z_h,\mu_b)\right)\, ,
\end{eqnarray}
with which the final result of the spin-dependent structure function is obtained as
\begin{align}
F_{UT}(Q;P_{hT})
=\frac{-1}{2z_h^3}\sum_q e_q^2\int_0^\infty  \frac{db~b^2}{(2\pi)}&J_1(P_{h T}/z_h~b)e^{-{S}_{\rm pert}(Q;b_*)-S_{\rm NP~Collins}^{\rm SIDIS}(Q;b)} \nonumber\\
&\left( \sum_i\delta C^{\rm (SIDIS)}_{q\gets i}\otimes h_{1}^{i/p}(x_B,\mu_b) \right)\left(\sum_j\delta \hat{C}_{j\gets q}^{\rm (SIDIS)}\otimes \hat H_{K/j}^{(3)}(z_h,\mu_b)\right) .
\label{fut_res}
\end{align}

\section{Numerical Estimate}

\label{sec:numerical}

Based on the formalism set up above, we will present the numerical estimate for the Collins asymmetry in Kaon production SIDIS process at the kinematics range of EicC and EIC in this section.

In order to obtain the numerical result of the unpolarized structure function in Eq.~(\ref{fuu_res}), we need to utilize the collinear unpolarized distribution function $f_1(x_B,\mu_b)$ and the collinear unpolarized fragmentation function $D_1(z_h,\mu_b)$ as the inputs of the evolution, for which we resort to the existed parametrizations.
For the collinear unpolarized distribution function $f_1(x_B)$ of the proton, we adopt the NLO set of the CT10 parametrization~\cite{Lai:2010vv}~(central PDF set), while for the fragmentation function $D_1(z_h)$ we apply the next-leading-order DSS parametrization\cite{deFlorian:2007aj}.
Besides, for $f_1$ and $D_1$, the free parameter $g_1$ in Eq.~(\ref{eq:g1}) is a very important ingredient in the evolution of the TMDs.
Here, we adopt $\langle k_\perp^2\rangle=0.57\rm{GeV}^2$, $\langle p_\perp^2\rangle=0.12\rm{GeV}^2$ given in Ref.~\cite{Anselmino:2013lza}. For the universal parameter $g_2$ in the nonperturbative Sudakov form factor, it has been extracted in the BLNY parametrization~\cite{Landry:2002ix} as $g_2 =0.184$.

\begin{figure}
  \centering
  \includegraphics[width=0.3\columnwidth]{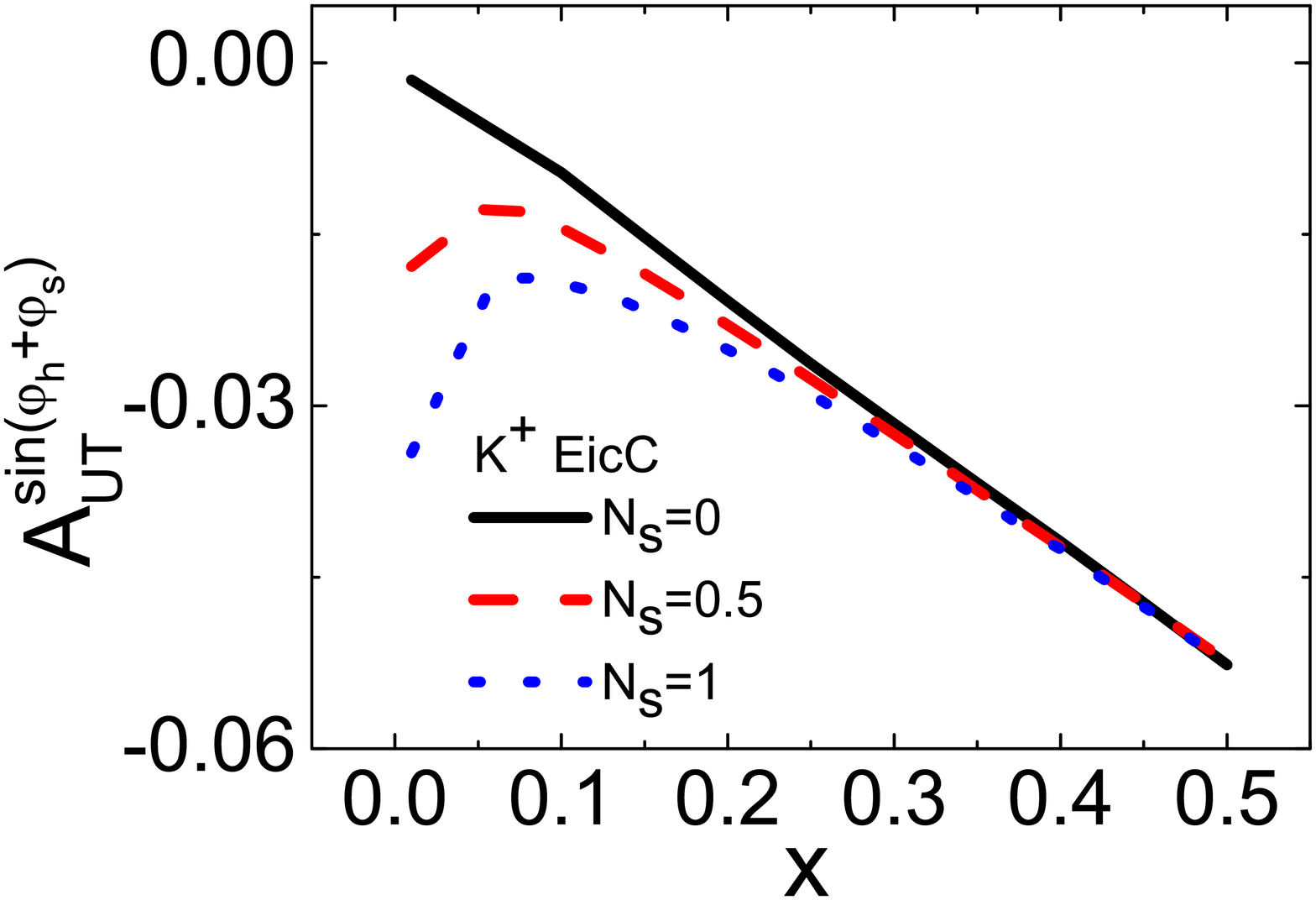}
  \includegraphics[width=0.3\columnwidth]{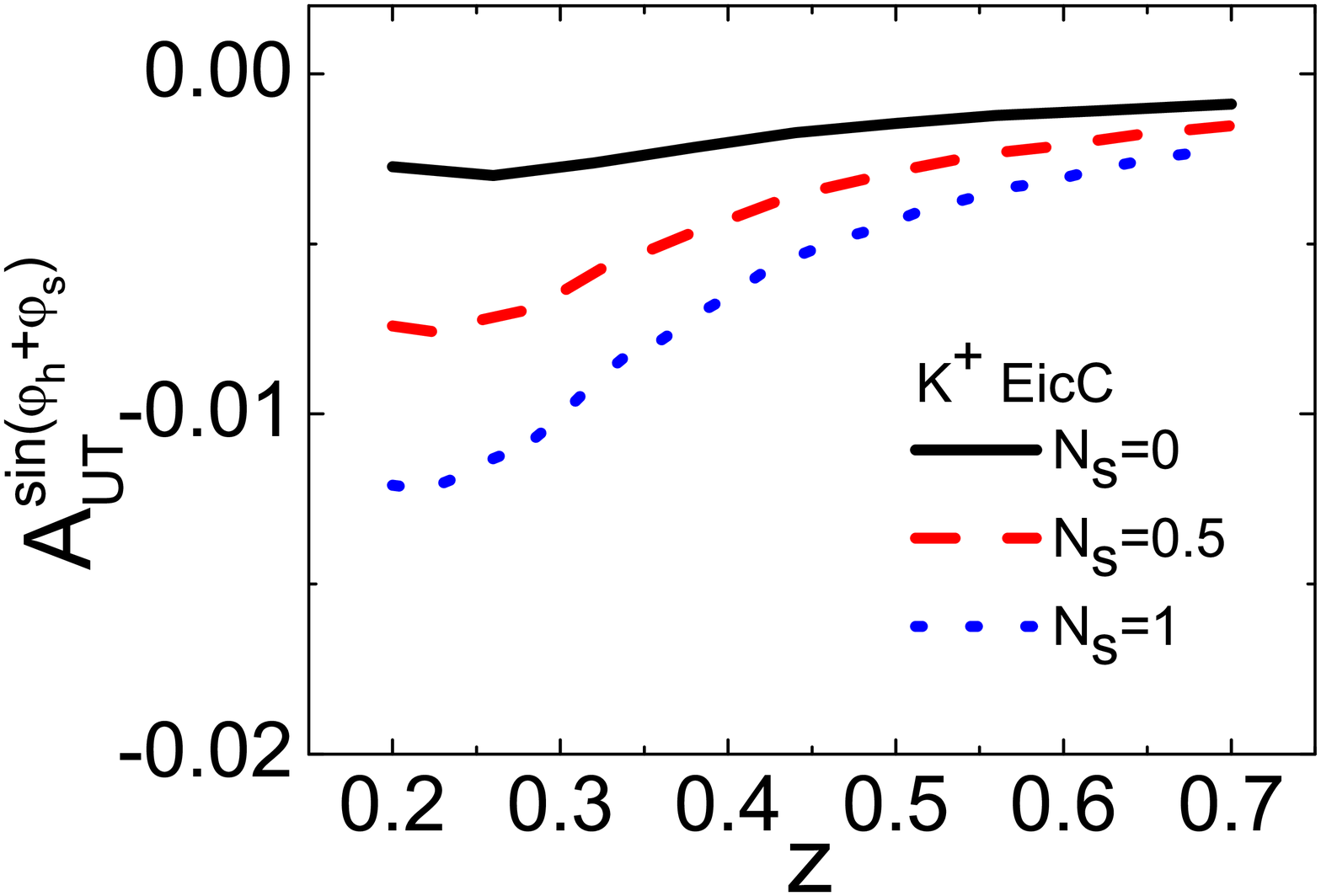}
  \includegraphics[width=0.3\columnwidth]{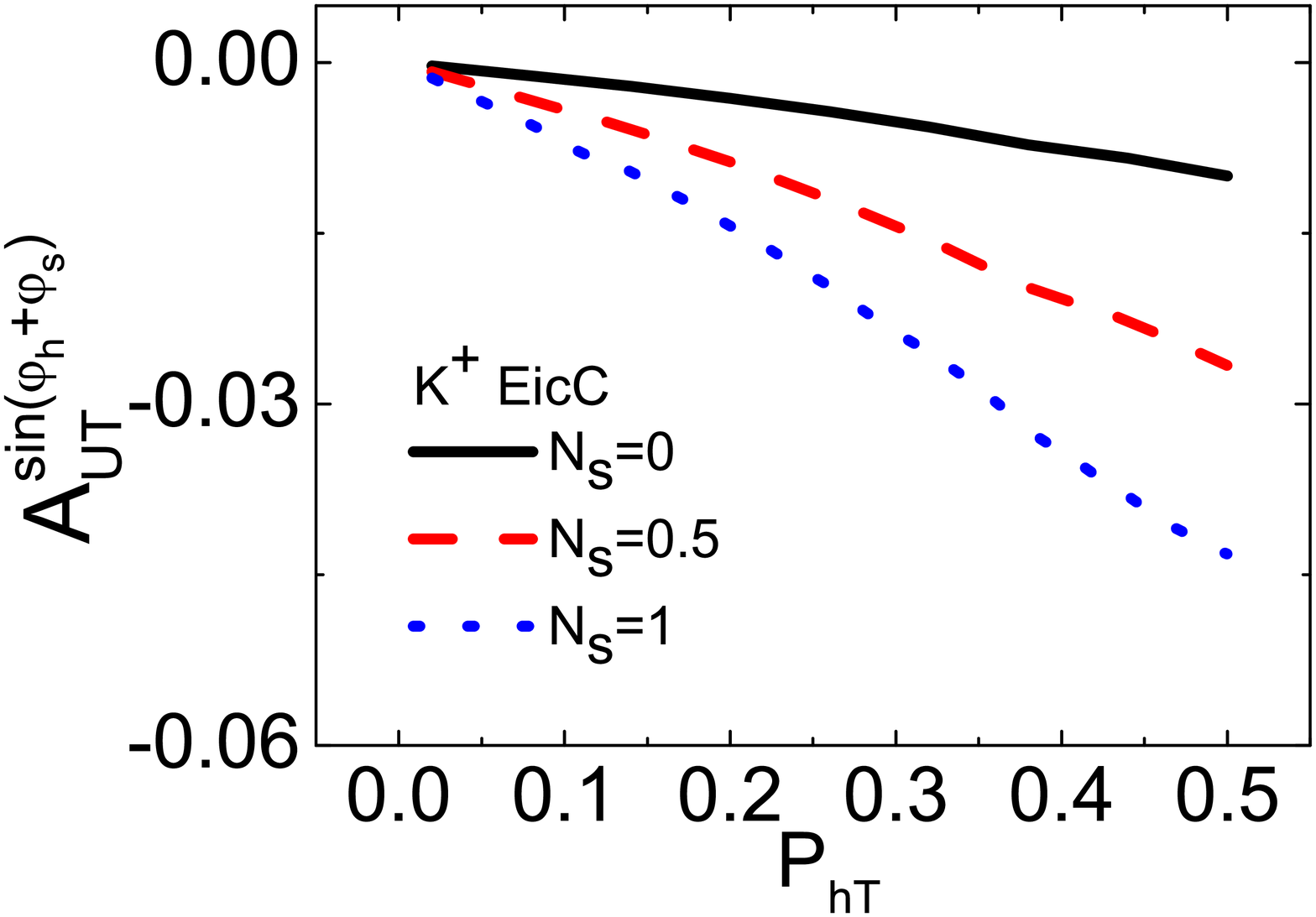}
  \includegraphics[width=0.3\columnwidth]{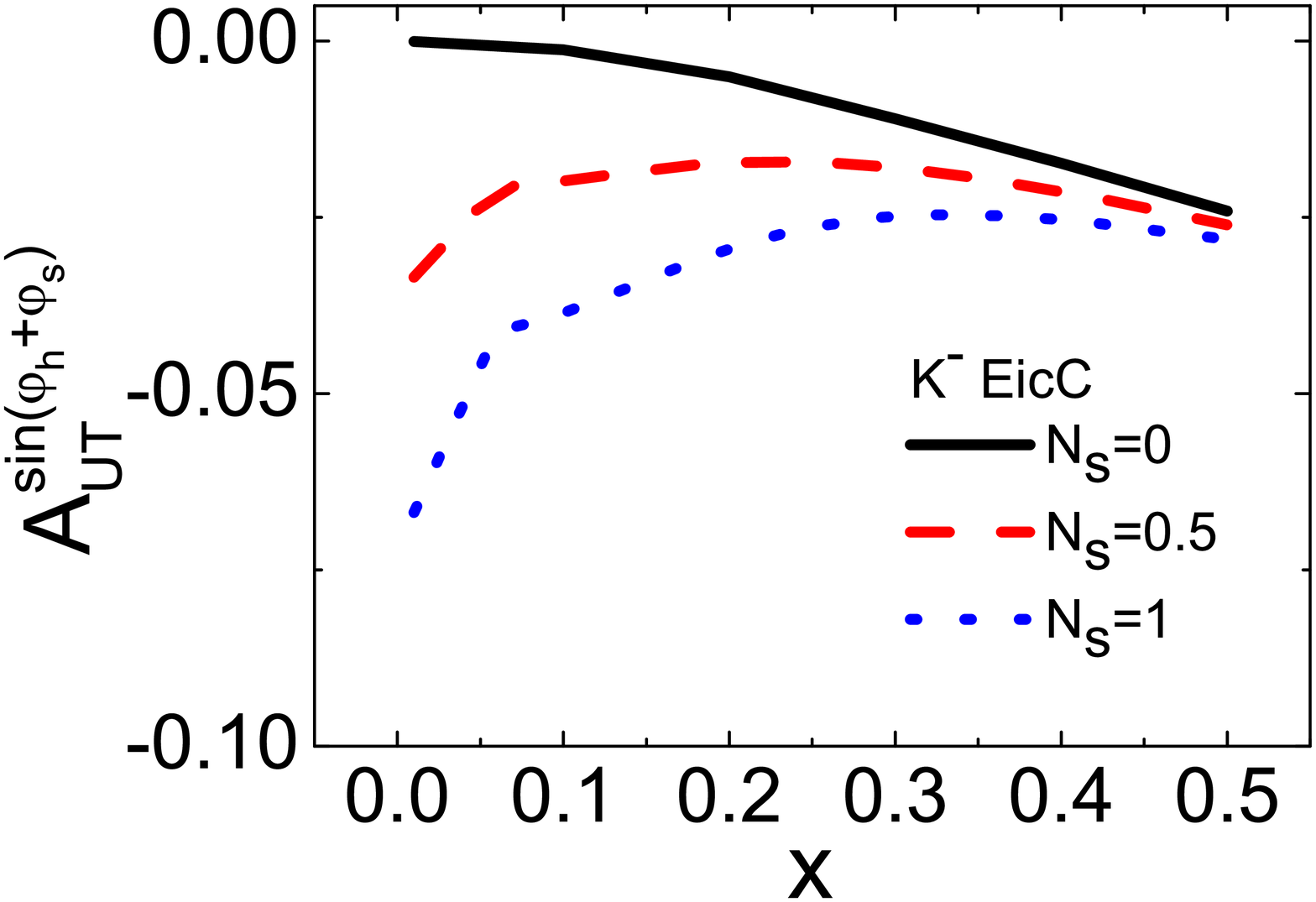}
  \includegraphics[width=0.3\columnwidth]{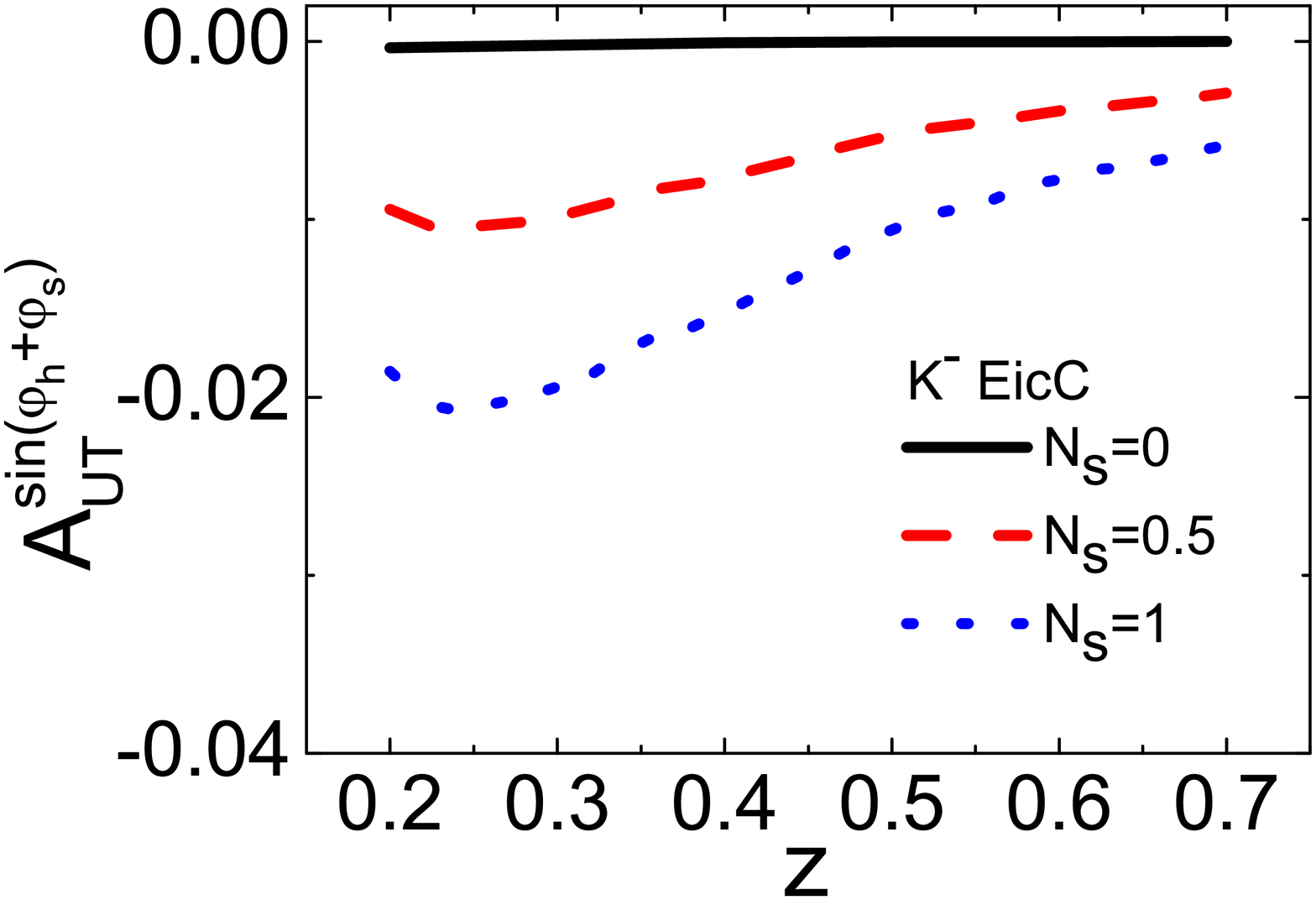}
  \includegraphics[width=0.3\columnwidth]{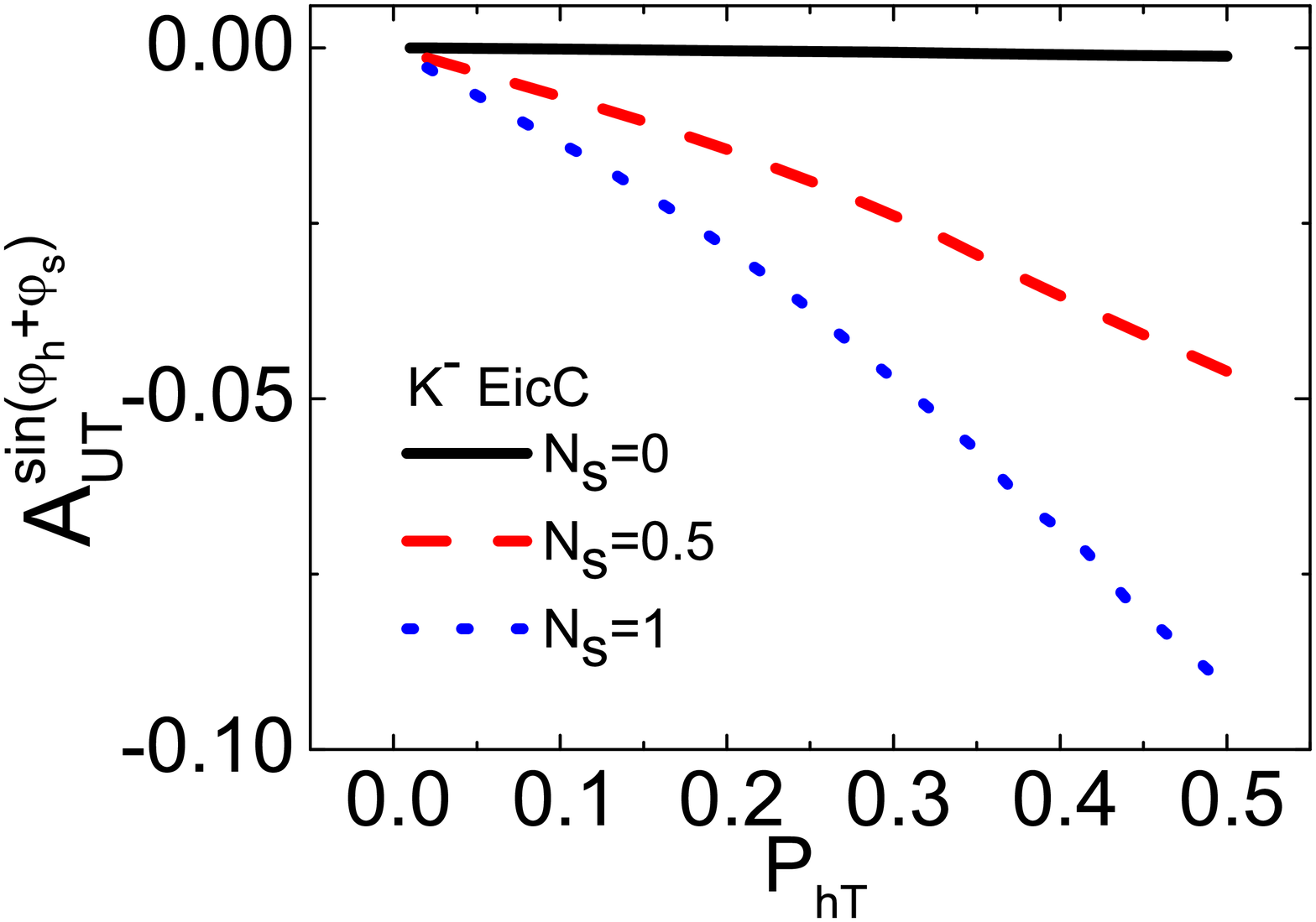}

  \includegraphics[width=0.3\columnwidth]{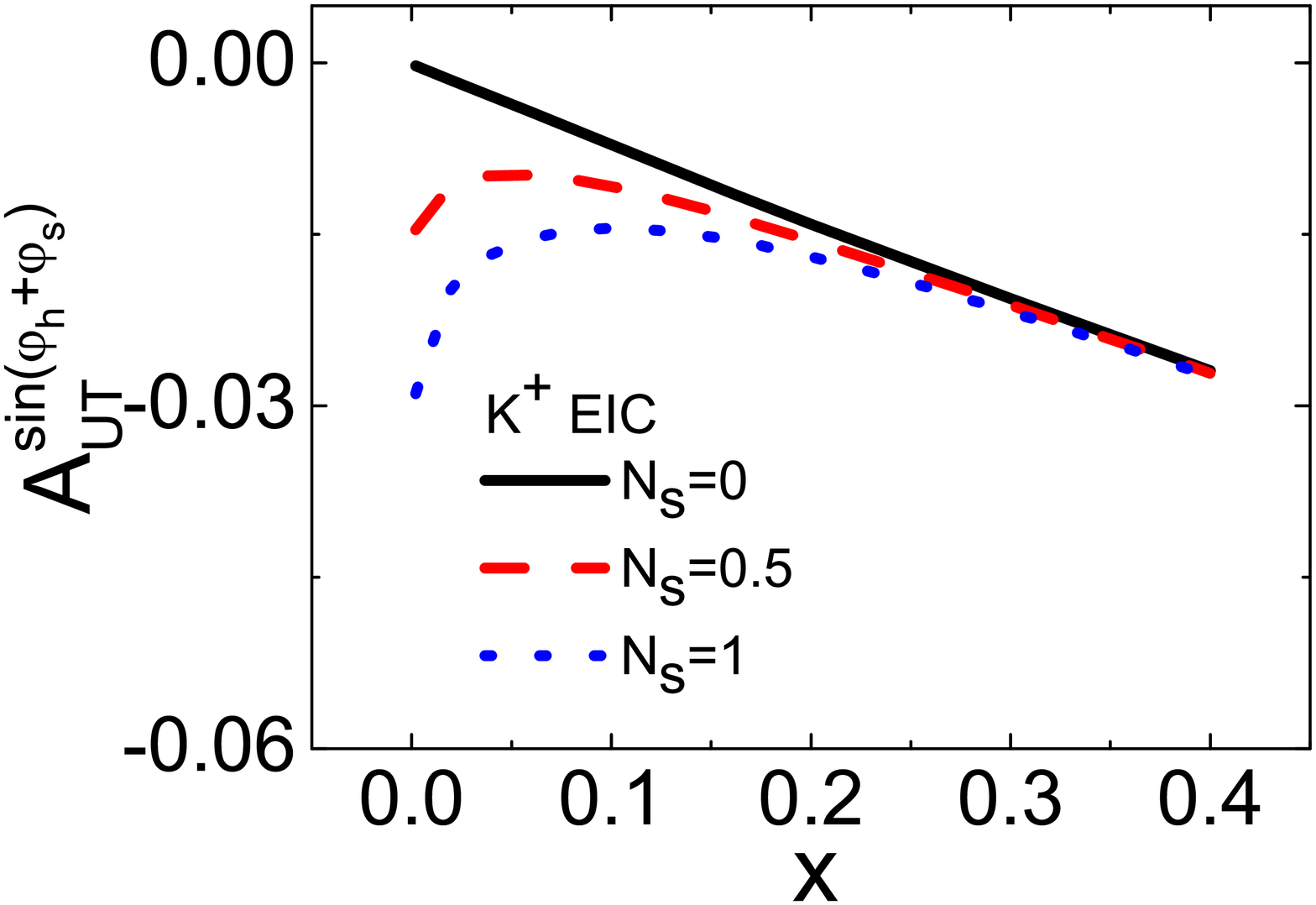}
  \includegraphics[width=0.3\columnwidth]{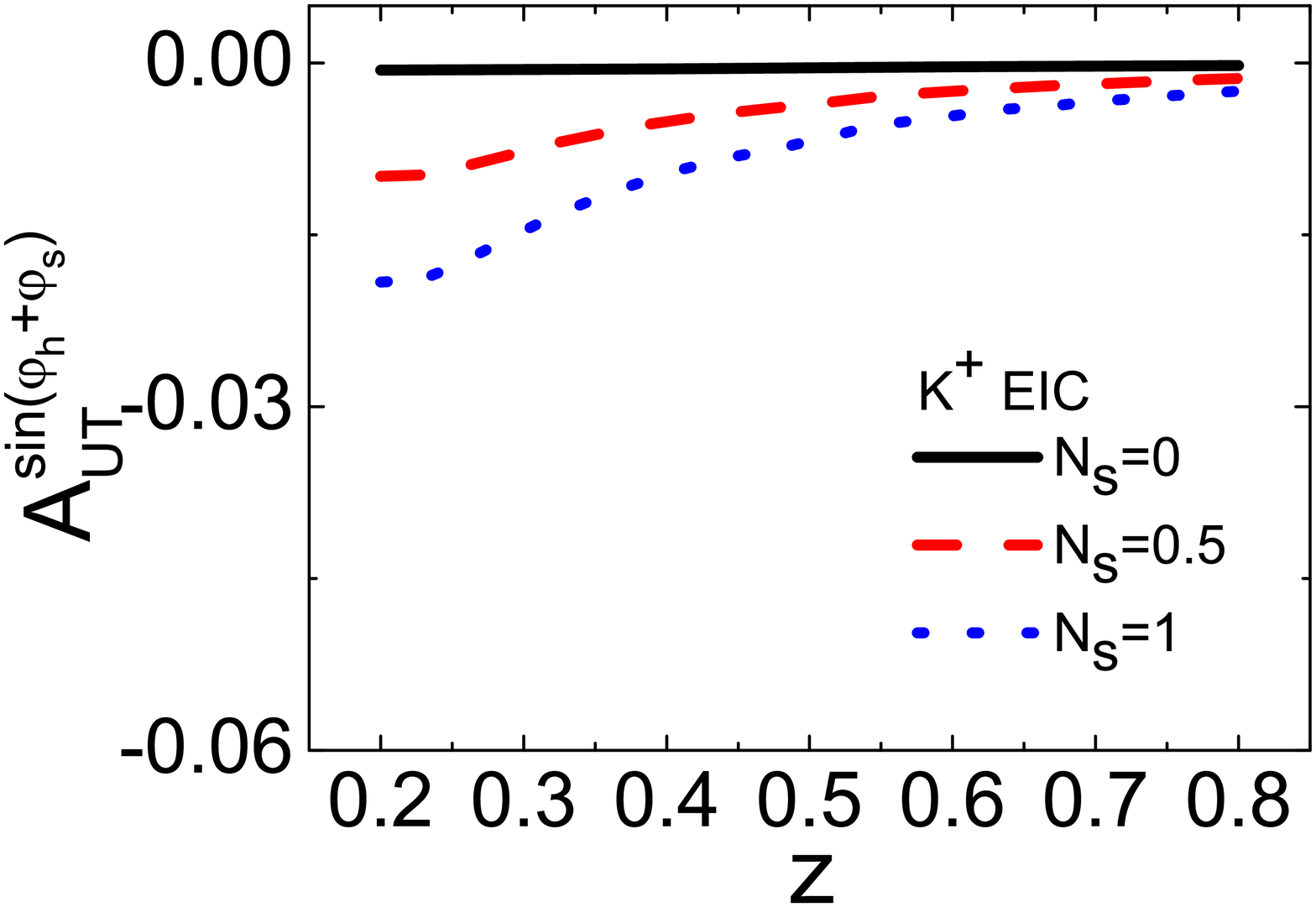}
  \includegraphics[width=0.3\columnwidth]{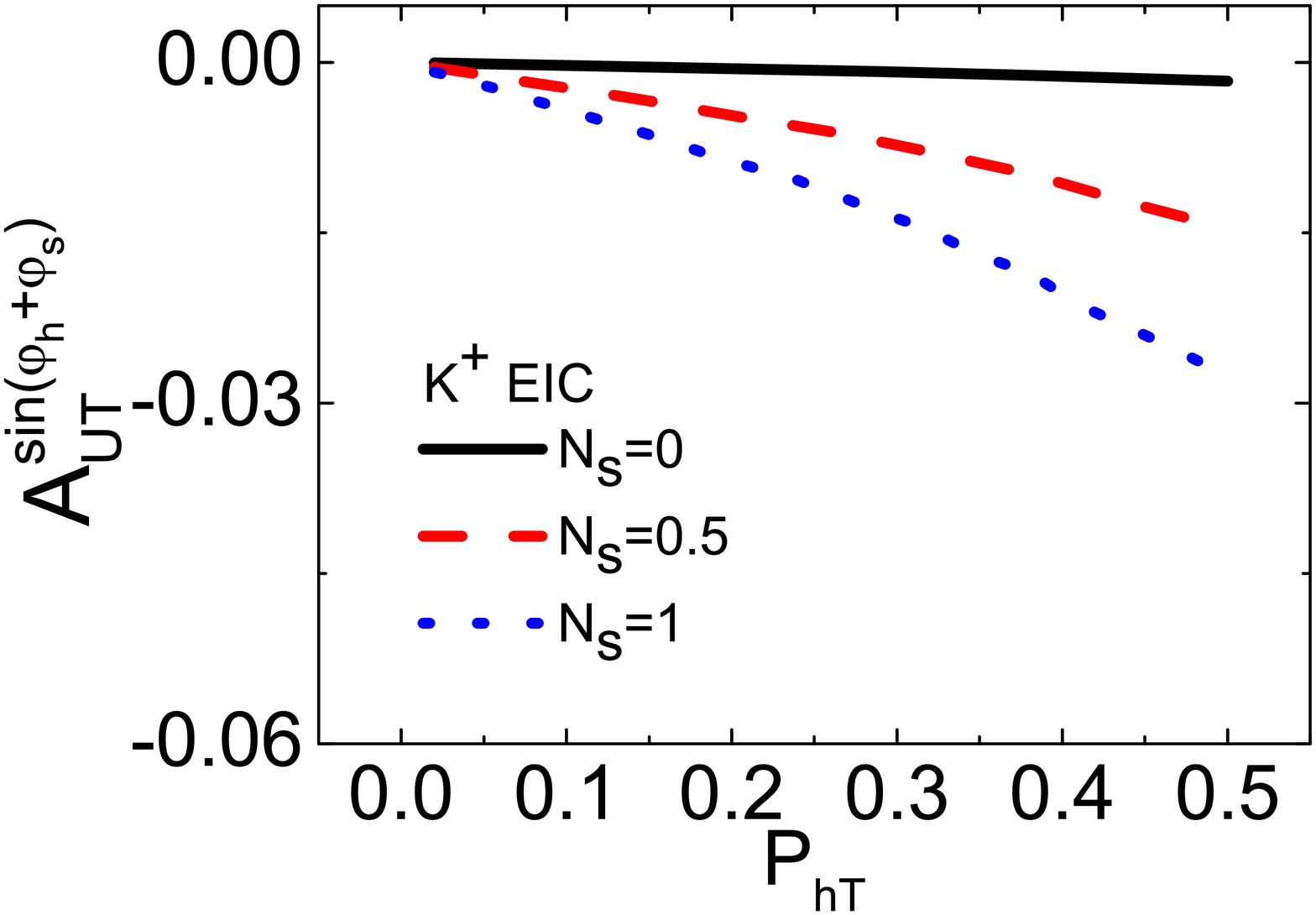}
  \includegraphics[width=0.3\columnwidth]{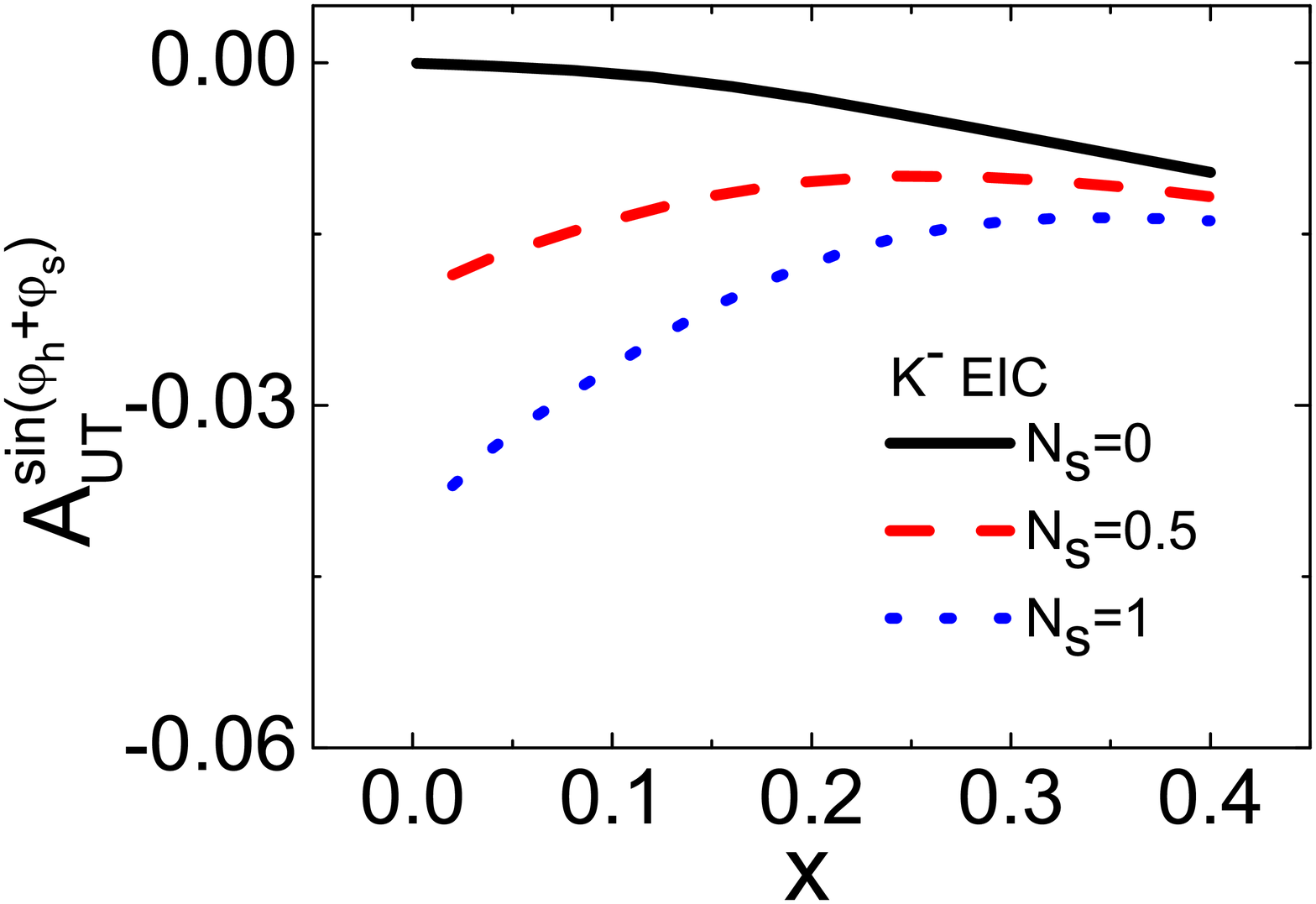}
  \includegraphics[width=0.3\columnwidth]{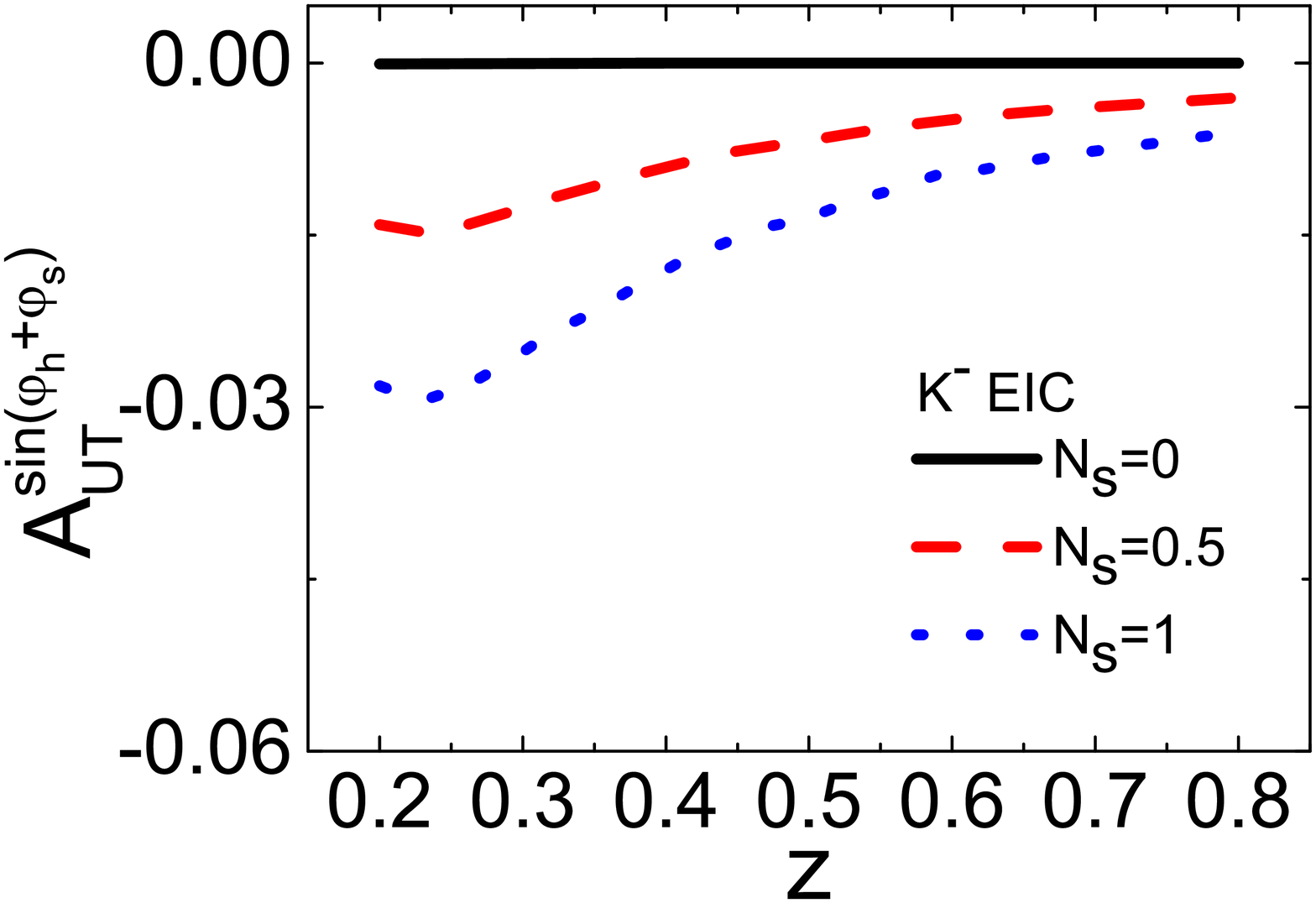}
  \includegraphics[width=0.3\columnwidth]{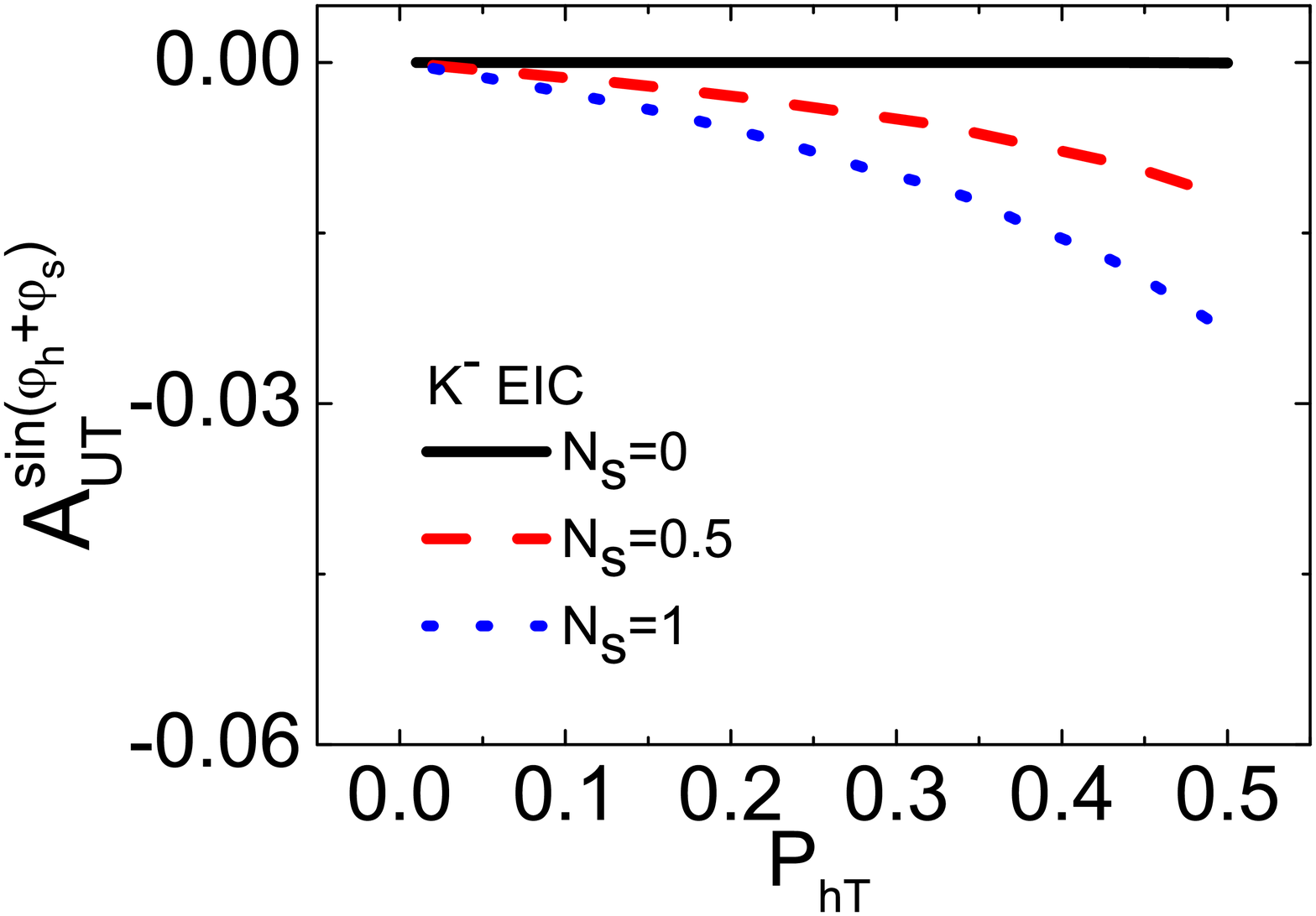}

  \caption{The Collins asymmetry in semi-inclusive Kaon production SIDIS at the kinematics of EIC and EicC as functions of $x$ (left panels), $z$ (middle panels), and $P_{hT}$ (right panels)}
  \label{fig:asy}
\end{figure}

In the case of the spin-dependent structure function in Eq.~(\ref{fut_res}), the collinear transversity function and the collinear correlation function of Collins function turn to be the inputs of the TMD evolution.
For the collinear transversity, we adopt the standard parameterization at the initial scale $Q^2_0=2.4~\rm{GeV}^2$ from Ref.~\cite{Kang:2015msa},
\begin{eqnarray}
h_1^{q}(x,Q_0) &=& N_{q}^h x^{a_{q}}(1-x)^{b_{q}} \frac{(a_{q} + b_{q})^{a_{q} + b_{q}}}{a_{q}^{a_{q}} b_{q}^{b_{q}}}  \frac{1}{2}(f_{1}^q(x,Q_0) + g_{1}^q(x,Q_0)   ) \ ,
\label{transversity}
\end{eqnarray}
where  $g_{1}^q$ is the helicity distribution function, for which we adopt the DSSV parametrization from Ref.~\cite{deFlorian:2009vb}.
For the $q=u$ and $d$ quarks, the parameterized results of $N_q^h$, $a_q$, and $b_q$   are taken from  the Table I of Ref.~\cite{Kang:2015msa}.
However, there is no information of the sea quark transversity distribution function, the high precision quantitative measurement of which is an important goal of the planned electron ion collider.
Thus, we assume the transversity of the sea quark at the initial energy scale has the form of
\begin{eqnarray}
h_1^{q}(x,Q_0) &=& N_s \frac{1}{2}(f_{1}^q(x,Q_0) + g_{1}^q(x,Q_0)   ) \ ,
\label{transversity_sea}
\end{eqnarray}
where $N_s\le 1$ is the factor taking into account the positivity bound.
To perform the DGLAP evolution of the transversity distribution function from the initial scale $Q_0^2 = 2.4\rm{ GeV}^2$ to the scale of $\mu_b=c_0/b_*$, the evolution package {\sc HOPPET}~\cite{Salam:2008qg} is applied and modified to include the evolution kernel of the transversity.
The value of the strong coupling $\alpha_s$ are consistently obtained at 2-loop order as
\begin{align}
\alpha_s(Q^2)&=\frac{12\pi}{(33-2n_f)\mathrm{ln}(Q^2/\Lambda^2_{QCD})}
\left\{{1-\frac{6(153-19n_f)}{(33-2n_f)^2}
\frac{\mathrm{ln}\mathrm{ln}(Q^2/\Lambda^2_{QCD})}{\mathrm{ln}(Q^2/\Lambda^2_{QCD})}}\right\} \label{eq:alphas}
\end{align}
with fixed $n_f=5$ and $\Lambda_{\mathrm{QCD}}=0.225\ \mathrm{GeV}$.
We note that the running coupling in Eq.~(\ref{eq:alphas}) satisfies $\alpha_s(M_Z^2)=0.118$ and the initial value $\alpha_s(Q_0) = 0.327$.

Although there is no direct parametrization for the collinear correlation function $\hat H_{h/q}^{(3)}(z_h)$, we can obtain it from the parametrization of the Collins function for the Kaon meson.
In Ref.~\cite{Anselmino:2015fty}, the Kaon Collins function is extracted from the
semi-inclusive hadron pair production in $e^+e^-$ annihilation, which turns out to be in good agreement with the measurements performed by the HERMES~\cite{Airapetian:2010ds} and COMPASS Collaborations\cite{Alekseev:2008aa,Adolph:2014zba}.
The Collins function was parameterized in Ref.~\cite{Anselmino:2015fty} as
\begin{align}
\Delta^N \! D_{h/q^\uparrow}(z_h,p_\perp) =  \tilde{\Delta} ^N D_{h/q^\uparrow}(z_h)
\> h(p_\perp)\,\frac{e^{-p_\perp^2/{\langle p_\perp^2\rangle}}}{\pi \langle p_\perp^2\rangle}\,.
\end{align}
Using the relation between $H_{1, h/q}^\perp(z_h,p_\perp)$ and $D_{h/q^\uparrow}(z_h,p_\perp)$:
\begin{align}
 H_{1\, h/j}^\perp(z_h,p_\perp)= \frac{z_h M_h}{2|p_\perp|} \Delta^N \! D_{h/q^\uparrow}(z_h,p_\perp) \;
\end{align}
and Eq.~(\ref{eq:H3}), one can obtain
the expression of $\hat H_{h/q}^{(3)}(z_h)$ as follows
\begin{align}
\hat{H}_{h/j}^{(3)}(z_h)= \frac{\sqrt{2e}}{M_C} \, {\cal N}^{C}_{q}(z_h)\, D_{h/q}(z_h)\, \left (\frac{M_C^2}{M_C^2+ \langle p_\perp^2\rangle} \right)^2  \langle p_\perp^2\rangle \; .
\end{align}
For the kaon Collins function, ${\cal N}^{C}_{q}(z_h)$ is set as the favored and disfavored constant in the parametrization~\cite{Anselmino:2015fty},
\begin{align}
{\cal N}^{C}_{q}(z_h)= N^{K}_{fav} =0.41^{+0.10}_{-0
.10} \; , \\
{\cal N}^{C}_{q}(z_h)= N^{K}_{dis} =0.08^{+0.18}_{-0.26} \; .
\end{align}
We should note that the Collins fragmentation functions of Kaon have huge errors due to the limitations of the experimental data available for Kaon production.
For the $g_1$ parameter needed in performing the TMD evolution of the transversity distribution function in Eq.~(\ref{eq:g1}), we assume the same Gaussian width of the transverse momentum value as $\langle k_\perp^2\rangle=0.57\rm{GeV}^2$~\cite{Anselmino:2015fty,Anselmino:2015sxa}. For the $g_1$ parameter of the Kaon Collins function in the non-perturbative Sudakov form factor, we adopt it from the parametrization in Ref.~\cite{Anselmino:2015sxa}
\begin{align}
g_1^{\rm H^{\perp}_1} = \frac{\langle p_\perp^2\rangle_{c}}{4z_h ^2}\ ,
\qquad
\langle p_\perp^2\rangle_{c}= \frac{M_C^2 \langle p_\perp^2\rangle}{M_C^2+ \langle p_\perp^2\rangle}\ ,
\end{align}
where $M_C^2=0.28  \,\rm{GeV}^2$~\cite{Anselmino:2015sxa}.
We shall note that the initial energy scale in the non-perturbative Sudakov form factor in Eq.~(\ref{eq:snp_gene}) is chosen as $Q^2_0=2.4\rm{GeV}^2$.
Also, the DGLAP evolution of $\hat H_{h/q}^{(3)}(z_h)$ was assumed to be the same as transversity, which means the homogenous term of the evolution kernel is adopted.

For the kinematical region that is available at EIC, our choices are as follows \cite{Accardi:2012qut}
\begin{align}
&0.001<x<0.4,\quad 0.07<y<0.9, \quad 0.2<z<0.8, \nonumber    \\
& 1\ \mathrm{GeV}^2<Q^{2} \;, \quad   W>5\ \mathrm{GeV},\quad \sqrt{s}=100 \ \mathrm{GeV},\quad P_{hT}<0.5 \ \mathrm{GeV}.
\end{align}
As for the EicC, we adopt the following kinematical cuts
\begin{align}
&0.005<x<0.5, \quad 0.07<y<0.9, \quad 0.2<z<0.7,  \nonumber    \\
 &1 \mathrm{GeV}^2<Q^2<200\ \mathrm{GeV}^2 \;, \quad   W>2\ \mathrm{GeV},\quad \sqrt{s}=16.7 \ \mathrm{GeV},\quad P_{hT}<0.5 \ \mathrm{GeV},
\end{align}
where $W^2=(P+q)^2\approx \frac{1-x}{x}Q^2$  is invariant mass of the virtual photon-nucleon system.
Since TMD factorization is proved to be valid to describe the physical observables in the region $P_{hT}\ll Q$, $P_{hT}<0.5 \ \mathrm{GeV}$ is chosen to guarantee the applicability of TMD factorization.
Using the above kinematical configurations and applying Eqs.~(\ref{eq:asymmetry}), (\ref{fuu_res}), and (\ref{fut_res}),
we calculate the transverse single-spin dependent Collins asymmetry of Kaon production in SIDIS process at EicC and EIC.
The corresponding numerical results are plotted in Fig.~\ref{fig:asy}, in which the left, middle, and right panels show the Collins asymmetry as functions of $x$, $z$, and $P_{hT}$, respectively.
The upper six panels show the predictions on Collins asymmetry at EicC for $K^+$ production and $K^-$ production; while the lower panels give the results at EIC for $K^+$ production and $K^-$ production.
In each figure, we plot the Collins asymmetry with the adjusting factor $N_s$=0~(no sea quark contribution in the initial energy scale of the transversity parametrization), 0.5, and 1 (saturate positivity bound) in the parametrization of the collinear transversity function for the sea quark using solid, dashed, and dotted lines, respectively.

As shown in Fig.~\ref{fig:asy}, the Collins asymmetry from our calculation is negative in all cases and sizable both at EicC and EIC.
The magnitude of $x$, $P_{hT}$ dependent asymmetries can reach around 0.05 at EicC and 0.03 at EIC.
For $z$ dependent asymmetry, the magnitude can reach 0.03 at EIC and 0.02 at EicC.
Our estimate also shows that the magnitude of the asymmetry increases  with increasing $N_s$ in Eq.~(\ref{transversity_sea}) of the collinear sea quark transversity function and the magnitudes at $N_s=1$ are about twice as large as those at $N_s=0.5$, which indicates that the sea quark transversity distribution function plays an important role in the Collins asymmetry of Kaon meson production.
The effect of the transversity of the sea quarks turns out to be smaller in the $K^+$ production process than that in the $K^-$ process.
The reason may be that the valence quarks for $K^+$ is $u\bar{s}$, for which Collins function needs to be convoluted with the transversity of the proton. There is only one sea quark transversity distribution function $h_1^{\bar{s}/p}$ coupled with the favored Collins function in the case.
While for the $K^-$ production process, the valence quarks of $K^-$ turn to be $\bar{u}s$, the relevant transversity distribution coupled to the favored Collins function becomes $h_1^{\bar{u}/p}$ and $h_1^{s/p}$, which are both sea quark transversity.
For the asymmetry as the function of $x$, there is a clear peak at $x\approx0.05$ at EicC
 when considering the non-zero sea quark transversity, while the peak vanishes with zero sea contribution of transversity.
Although the peak turns to be vague at EIC, the tendency still remains.
The $z$, $P_{hT}$ dependent asymmetries are rather small in the case of vanishing sea quark transversity.
Thus, there is a great opportunity to access the sea quark transversity by utilizing the electron ion colliders to measure the Collins asymmetry of Kaon production in SIDIS process.
However, due to the limited amount of the experimental data, there are large errors in the parametrization of the Collins function for the Kaon meson, of which the knowledge is relatively limited. Thus, we also emphasize the importance of the $e^+e^-$ data in the extraction of the Collins function for the Kaon meson. Combining the experimental data from SIDIS process and those from $e^+e^-$ process, one can perform the global analysis of data sets and simultaneously extract the Kaon Collins function as well as the transversity distribution function for both valence quark and sea quark.

\section{Conclusion}
\label{sec:conclusion}

In this work, we have applied the TMD factorization formalism to study the single transverse-spin Collins asymmetry $\sin\left(\phi_h +\phi_s\right)$ modulation of Kaon production in SIDIS process at the kinematics configurations of EIC and EicC.
The asymmetry arises from the coupling of the target proton transversity and the Collins fragmentation function of the Kaon meson.
We have taken into account the TMD evolution of distributions and fragmentation functions by including the Sudakov form factors.
For the nonperturbative Sudakov form factor associated with the TMDs, we have adopted the traditional Gaussian form.
The hard coefficients associated with the corresponding collinear functions are kept in the next-to-leading-logarithmic order.
For the transversity distribution of the proton needed, we have employed a recent parametrization for which the TMD evolution effect is considered.
For the Collins function of the producing Kaon, we adopted the available parametrization, which can well describe the SIDIS data.
Our results demonstrated that,
the Collins asymmetry of Kaon production in SIDIS process is sizable at the kinematics configurations of both EicC and EIC.Thereby it could be measured by these experiments in the future.
Furthermore, we have considered the contribution of sea quark transversity to the asymmetry.
The numerical calculation showed that different choices of the sea quark transversity ($N_s=0$, 0.5, and 1) lead to different sizes of the asymmetry, particularly in the case of $K^-$ production.
Therefore, the measurement on the Collins asymmetry of semi-inclusive Kaon production at future electron ion colliders can provide useful constraints on the sea quark transversity.
We also note that there are large errors in the extraction of the Kaon Collins function, which indicates the importance of more precision $e^+e^-$ data in order to constrain the Kaon Collins function.

\section{Acknowledgments}
This work is partially supported by the NSFC (China) grants 11905187, 11847217,11575043 and 11120101004.
X. Wang is supported by the China Postdoctoral Science Foundation under Grant No.~2018M640680.

\end{document}